\documentclass[12pt]{iopart}
\usepackage{iopams}
\usepackage{graphicx}

\newcommand{\fm}{\;\mbox{fm}}
\newcommand{\MeV}{\;\mbox{MeV}}
\newcommand{\GeV}{\;\mbox{GeV}}
\newcommand{\Nc}{N_{\rm c}}
\newcommand{\Nf}{N_{\rm f}}
\newcommand{\Nt}{N_\tau}
\newcommand{\Tc}{T_{\rm c}}
\newcommand{\Tf}{T_{\rm f}}
\newcommand{\LQCD}{\Lambda_{\rm QCD}}
\newcommand{\diag}{\mathop{\rm diag}}
\newcommand{\as}{\alpha_{\rm s}}

\newcommand{\qmod}{q_{\rm mod1}}
\newcommand{\muB}{\mu_{\rm B}}
\newcommand{\muq}{\mu_{\rm q}}
\newcommand{\muf}{\mu_{\rm f}}
\newcommand{\muI}{\mu_{\rm I}}
\newcommand{\muIm}{\tilde{\mu}_{\rm q}}
\newcommand{\mq}{m_{\rm q}}
\newcommand{\Mq}{M_{\rm q}}

\newcommand{\calP}{\mathcal{P}}
\newcommand{\calCP}{\mathcal{CP}}
\newcommand{\bE}{\bi{E}}
\newcommand{\bB}{\bi{B}}
\newcommand{\bx}{\bi{x}}

\newcommand{\bj}{\bi{j}}
\newcommand{\bJ}{\bi{J}}
\newcommand{\Md}{\mathcal{M}_{\rm quark}}

\begin{document}
\jl{4}
\topical{QCD matter in extreme environments}
\author{K Fukushima}
\address{Department of Physics, Keio University,\\
         3-14-1 Hiyoshi, Kohoku-ku, Yokohama-shi, Kanagawa 223-8522, Japan}

\begin{abstract}
We review various theoretical approaches to the states of QCD matter
out of quarks and gluons in extreme environments such as the
high-temperature states at zero and finite baryon density and the
dimensionally reduced state under an intense magnetic field.  The
topics at high temperature include the Polyakov loop and the 't~Hooft
loop in the perturbative regime, the Polyakov loop behaviour and the
phase transition in some of non-perturbative methods;  the
strong-coupling expansion, the large-$\Nc$ limit and the holographic
QCD models.  These analyses are extended to hot and dense matter with
a finite baryon chemical potential.  We point out that the difficulty
in the finite-density problem has similarity to that under a strong
magnetic field.  We make a brief summary of results related to the
topological contents probed by the magnetic field and the Chiral
Magnetic Effect.  We also address the close connection to the (1+1)
dimensional system.
\end{abstract}
\submitted



\section{Introduction}

The inside of heavy nuclei is already a very interesting and peculiar
environment.  It is known that the density of nucleons, i.e.\ the
baryon density, takes an almost constant value,
$\rho_0\simeq 0.17\,\mbox{nucleon}/\mbox{fm}^3$, in the central region
of nuclei independently of the atomic number $A$ for large enough
$A$.  In terms of our daily units this normal nuclear density is as
huge as $\sim 10^{12}\,\mbox{g}/\mbox{cm}^3$.  When two heavy nuclei
(positively charged ions) collide at almost the speed of light, an
enormous energy is crammed in a volume of size of heavy nuclei with
the (transverse) radius $r_A\sim 1.2A^{1/3}\fm$.  In this way the
relativistic heavy-ion collision experiment provides us with an ideal
opportunity to examine the state of matter under extreme environments
as have ever existed in the Universe.

The collision energy is finally released into a form of the energy
conveyed by produced particles.  Measuring the momentum distribution
of those particles, the initial energy density can be deduced, which
is summed up as the Bjorken formula (though it looks slightly
different from the original form \cite{Bjorken:1982qr}),
\begin{equation}
 \epsilon_0 = \frac{\langle m_\perp\rangle}{\tau_0 \pi r_A^2} \cdot
            \frac{\rmd N}{\rmd y} \;,
\end{equation}
where $\tau=\sqrt{t^2-z^2}$ and $y=\frac{1}{2}\ln[(t+z)/(t-z)]$ are
the proper time and the coordinate rapidity, respectively.  Each
produced particle has energy $m_\perp=\sqrt{p_\perp^2+m^2}$ with the
transverse momentum $p_\perp$.  The number of particle is represented
by $N$.  The energy density is then given by the energy of particle
$\langle m_\perp\rangle\,\rmd N$ divided by the initial volume
$\tau_0 \pi r_A^2 \rmd y$, where $\tau_0$ denotes the initial time
when particles are produced.

Once the thermal equilibrium is reached and deconfined gluons and
$\Nf$-flavour quarks -- the fundamental objects in Quantum
Chromodynamics (QCD) -- are assumed to be massless particles, the
energy density $\epsilon$ is translated into the temperature $T$
through the Stefan-Boltzmann law; $\epsilon = (\pi^2/30)\, n(T) T^4$
where $n(T) = 16~\mbox{(gluons)} + 10.5\Nf~\mbox{(quarks)}$ is the
effective number of physical degrees of freedom.  The Monte-Carlo
simulation of finite-$T$ QCD discretized on the lattice (see
references \cite{Borsanyi:2010bp,Bazavov:2010bx} for recent advances)
has identified the pseudo-critical temperature as
$\Tc\simeq (150$--$160)\MeV$.  This value of $\Tc$ corresponds to the
critical energy density, $\epsilon_{\rm c}\simeq
(0.8$--$1.0)\GeV/\mbox{fm}^3$, beyond which the state of matter should
be composed of gluons and quarks, namely, a quark-gluon plasma (QGP)
is realised.  We note that the energy density at normal nuclear
density $\rho_0$ is $m_N\rho_0\simeq 0.16\GeV/\mbox{fm}^3$ with the
nucleon mass $m_N=0.94\GeV$ used, that is about one fifth smaller than
$\epsilon_{\rm c}$.

In the facilities called the Alternating Gradient Synchrotron (AGS) at
BNL and the Super Proton Synchrotron (SPS) at CERN the heavy-ion
(Au-Au at AGS and Pb-Pb at SPS) experiments with a fixed target had
been conducted since 1986.  There, at the collision energy (per
nucleon-nucleon) $\sqrt{s_{_{NN}}}=17.2\GeV$ at SPS, some theoretical
studies \cite{Vitev:2004bh} led to an estimate
$\epsilon_0=(1.2$--$2.6)\GeV/\mbox{fm}^3$ (at $\tau_0=0.8\fm$).  This
initial energy density exceeds the critical value, so that the state
of matter created at SPS could be a QGP possibly.  It has become
evident in the Relativistic Heavy Ion Collider (RHIC) at BNL, which
has been in operation since 2000, that the collision energy
$\sqrt{s_{_{NN}}}=200\GeV$ is high enough to form the QGP with the
initial energy density $\epsilon_0=(12$--$20)\GeV/\mbox{fm}^3$ (at
$\tau_0=0.6\fm$).

The experimental activities are still continued;  there are two major
directions as the future plan;  one is the direction towards higher
$T$ at larger collision energies, while the other is the direction
towards higher baryon density at smaller energies.  The former is
already ongoing since 2010 in the Large Hadron Collider (LHC) at CERN,
where the experimental data from the Pb-Pb collision at
$\sqrt{s_{_{NN}}}=2.76\;\mbox{TeV}$ are significantly improving the
quality of analysis.  It is said that the QGP physics has been
promoted from the ``discovery stage'' to the ``precision science''
studies.  The latter direction, i.e.\ the extrapolation to higher
baryon density regions and the experimental survey over the whole QCD
phase diagram, is also underway.  The beam-energy scan program at RHIC
aims to give a detailed portrait of the phase structure of QCD matter,
especially to locate a special point of the exact second-order phase
transition called the QCD critical point
\cite{Asakawa:1989bq,Barducci:1989wi,Berges:1998rc,Stephanov:1998dy,%
Stephanov:1999zu}, which would serve as a landmark.  The RHIC energy
scan will be complemented by future experiments planned in the
Facility for Antiproton and Ion Research (FAIR) at GSI, the
Nuclotron-based Ion Collider Facility (NICA) at JINR and perhaps the
Japan Proton Accelerator Research Complex (JPARC) at JAEA and KEK.

This review is a self-contained summary of some selected approaches in
theory to the physical properties of QCD matter in extreme
environments such as the high temperature in \sref{sec:temperature}
and the finite baryon density in \sref{sec:density}.  It is, however,
practically impossible to cover all the topics related to finite
temperature/density QCD here.  We shall specifically focus on the
deconfinement physics and the dynamics of the order parameter called
the Polyakov loop.  As for the phase structure associated with
chiral-symmetry breaking and restoration, the interested readers may
consult my reviews \cite{Fukushima:2008pe,Fukushima:2010bq} and other
reviews
\cite{Gross:1980br,Svetitsky:1985ye,Klevansky:1992qe,Hatsuda:1994pi,%
MeyerOrtmanns:1996ea,Smilga:1996cm,Rischke:2003mt} and references
therein.  In particular my previous reviews
\cite{Fukushima:2008pe,Fukushima:2010bq} include a pedagogical
introduction to more generic physics of hot and dense QCD and subjects
related to chiral-symmetry breaking, but not much about deconfinement
physics.  Hence, this present review is complementary to
\cite{Fukushima:2008pe,Fukushima:2010bq} in respect to deconfinement
physics and the Polyakov loop dynamics.

The last half of this review is devoted to a new physics possibility
in a strong magnetic field $\bB$ produced by non-central collisions.
The strength of this produced $\bB$ surpasses the surface magnetic
fields on the neutron star by orders of the magnitude.  The presence
of the magnetic background which is as strong as the QCD energy scale
$\LQCD\sim200\MeV$ may well affect experimental observables
considerably.  A systematic procedure to resum relevant diagrams
important at strong $\bB$ has not been better established on the
practical level than the finite temperature/density field theory.
Therefore our discussions in \sref{sec:magnetic} shall not be
conclusive enough but they will aim to be comprehensive over various
aspects of the phenomenon known as the Chiral Magnetic Effect.


\section{High-temperature State of QCD}
\label{sec:temperature}

At sufficiently high temperature we can make use of the perturbative
expansion in terms of the strong coupling constant, $g$, since the
renormalisation of the UV divergences and the independence of the
renormalisation point make $g$ or $\as=g^2/4\pi$ run as a function of
the momentum scale $\mu$;  at the one-loop order,
\begin{equation}
 \as(\mu^2) = \frac{\as(\mu_0^2)}{1+\beta_0\, \as(\mu_0^2)
  \ln(\mu^2/\mu_0^2)} \;,
\label{eq:running}
\end{equation}
where $\beta_0=(11-\frac{2}{3}\Nf)/4\pi$.  For the reference scale
$\mu_0$, it is a conventional choice to set $\mu_0$ as the $Z^0$ boson
mass, $M_Z=91.2\GeV$ and the world average at present is
$\as(M_Z^2) = 0.1184(7)$ \cite{Nakamura:2010zzi}.  It is natural from
equation~\eref{eq:running} to consider that the strong coupling
constant gets smaller at higher temperature where the typical momentum
scale among thermally interacting particles is characterised by the
temperature $T$.  One could therefore take $\mu\propto T$ but cannot
precisely fix the proportionality coefficient because it is not clear
which renormalisation condition is the most efficient to resum the
higher-order diagrams.  The conventional prescription is to take
$\mu=2\pi T$ and vary $\mu$ to check the stability of the physical
results \cite{Andersen:2002ey} (see also \cite{Andersen:2004fp} for a
review).

While the perturbative QCD calculations are useful at high temperature
(for the state-of-the-art calculations up to three-loop order, see
\cite{Andersen:2010ct,Andersen:2011sf}), it is necessary to develop
a non-perturbative method to go down towards $\Tc$.  In this section
we first look over the perturbative results, and then, we will proceed
to several non-perturbative approaches.  As we have stated, in this
review, we mostly address the pure gluonic sector and the
deconfinement order parameter.  We postpone the discussions on
dynamical quark effects to \sref{sec:density}.

\begin{figure}
\begin{minipage}{0.4\textwidth}
 \begin{center} (a) \\
  \includegraphics[width=0.7\textwidth]{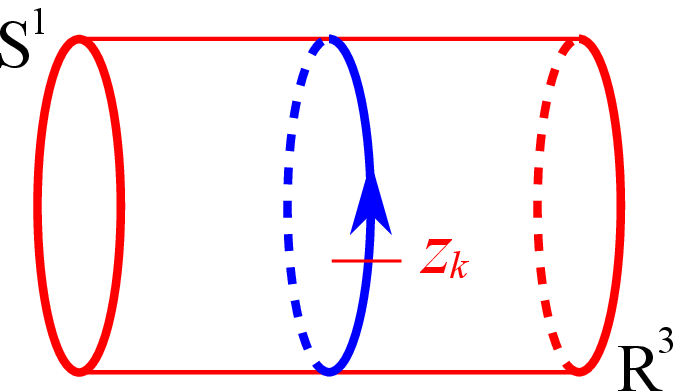}
 \end{center}
\end{minipage}
\begin{minipage}{0.6\textwidth}
 \begin{center} (b) \\
 \includegraphics[width=0.94\textwidth]{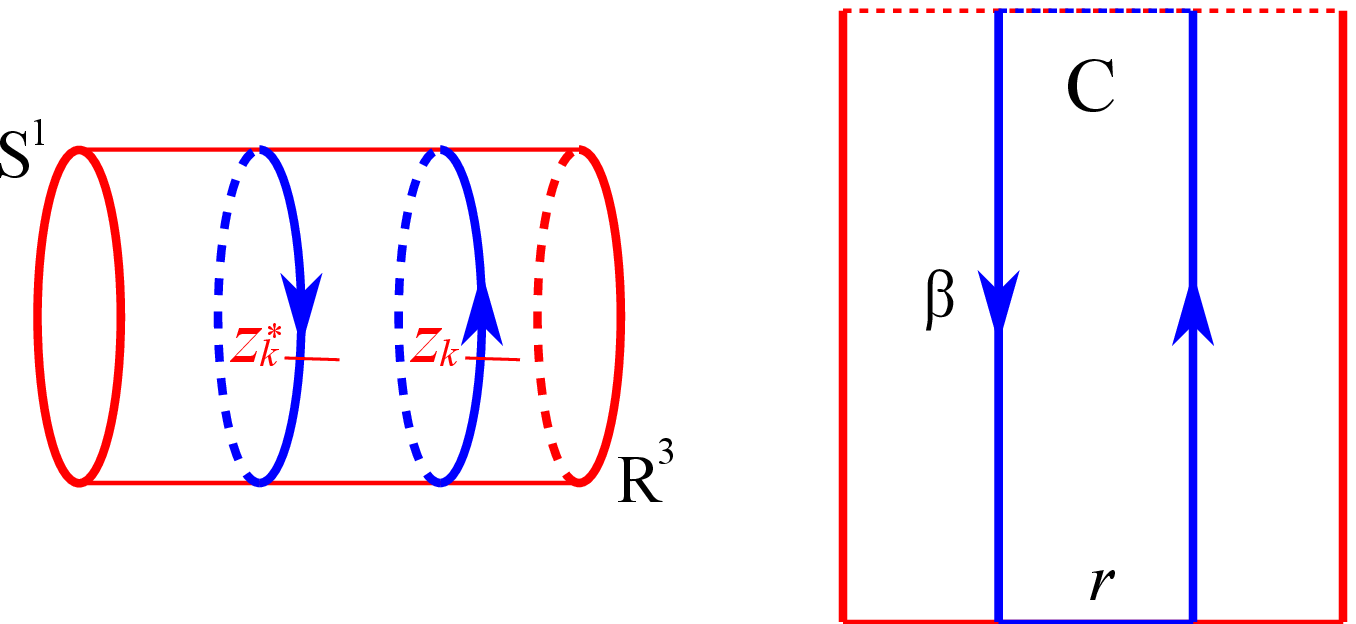}
 \end{center}
\end{minipage}
\caption{(a) Graphical representation of the Polyakov loop in
  Euclidean space-time.  The centre transformation multiplies $z_k$ on
  the Polyakov loop matrix.\ \ (b) Left: Polyakov loop correlator that
  is a counterpart of the closed Wilson loop.  The correlator is
  centre invariant because $z_k\cdot z_k^\ast=1$.\ \ Right: An
  opened-up figure whose temporal edges are contracted by the periodic
  boundary condition.}
\label{fig:pol}
\end{figure}

The conventional choice of the order parameter for quark deconfinement
at finite temperature is the (traced) Polyakov loop
\cite{Polyakov:1978vu,Susskind:1979up}.  The Polyakov loop matrix and
its traced quantity are denoted respectively as
\begin{equation}
 L = \mathcal{P}\exp\biggl[ \rmi g\int_0^\beta \rmd x_4\,
  A_4(\bx,x_4) \biggr] \;,\qquad
 \Phi = \frac{1}{\Nc}\langle \tr L \rangle \;,
\label{eq:pol}
\end{equation}
in the imaginary-time formalism of the finite-temperature field
theory.  \Fref{fig:pol}~(a) is a graphical representation on the
manifold of $\mathbb{S}^1\times\mathbb{R}^3$.  The Polyakov loop
expectation value, $\Phi$, can be interpreted as the partition
function in the presence of a static-quark source.  The logarithm of
$\Phi$ thus yields the single-quark free energy $f_{\rm q}$ as in
standard thermodynamics; $f_{\rm q}=-T\ln\Phi$.  In the quark
deconfined phase $\Phi$ and $f_{\rm q}$ take a finite value, whereas
$\Phi\to0$ and $f_{\rm q}\to\infty$ in the quark confined phase.

This behaviour of the Polyakov loop is understood from Wilson's
standard criterion of confinement \cite{Wilson:1974sk}.  As seen in
\fref{fig:pol}~(b) the Wilson loop on $\mathbb{S}^1\times\mathbb{R}^3$
amounts to the correlator of the Polyakov loop $L$ and the
anti-Polyakov loop $L^\dagger$.  In the confined phase the Wilson loop
shows the area law (in the absence of dynamical quarks) and the
deconfinement phase results in the perimeter law, which can be
interpreted in terms of the Polyakov loop as
\begin{eqnarray}
\hspace{-3em}
 & \mbox{\underline{Confined Phase}} \nonumber\\
\hspace{-3em}
 & \quad \langle W(C)\rangle \sim \rme^{-\sigma_{\rm w} \Sigma(C)}
  \;\Rightarrow\; \langle\tr L^\dagger(r\to\infty)\tr L(0) \rangle \to 0
  \qquad\:\:\:\:\Rightarrow\; \Phi = 0 \\
\hspace{-3em}
 & \mbox{\underline{Deconfined Phase}} \nonumber\\
\hspace{-3em}
 & \quad \langle W(C)\rangle \sim \rme^{-\sigma_{\rm w}' L(C)}
  \;\Rightarrow\; \langle\tr L^\dagger(r\to\infty)\tr L(0) \rangle \to
   \mbox{(const.)} \;\Rightarrow\; \Phi \neq 0 \;,
\end{eqnarray}
where $\Sigma(C)=\beta r$ is the area enclosed by $C$ and $L(C)$ is
the perimeter (see \fref{fig:pol}~(b)).

This criterion is related to the global symmetry of QCD that makes the
Polyakov loop expectation value vanishing.  To manifest this symmetry
let us turn to the lattice approximation to equation~\eref{eq:pol},
i.e.
\begin{equation}
 L = \prod_{n_4=0}^{\Nt-1} U_4(\bi{x},x_4=a n_4) \;.
\label{eq:pol-lat}
\end{equation}
Here $U_4=\rme^{-\rmi g a A_4}$ is the temporal link variable.  (Note
that in equation~\eref{eq:pol-lat} the time product is from the left
to the right with increasing time, while later time comes to the
left in equation~\eref{eq:pol}.)  Then, under a general gauge
transformation, the link variable changes as
\begin{equation}
 U_\mu(x) \;\to\; V(x) U_\mu(x) V^\dagger(x+a\hat{\mu}) \;.
\end{equation}
Therefore, the following transformation belongs to a subgroup of the
gauge transformation,
\begin{equation}
 U_4(\bx,x_4=a(N_\tau-1)) \;\to\; z_k \, U_4(\bx,x_4=a(N_\tau-1))
\label{eq:center-trans}
\end{equation}
for all $\bx$ with an $\Nc\times \Nc$ matrix,
$z_k=\diag(\rme^{2\pi\rmi k/\Nc},\rme^{2\pi\rmi k/\Nc},\dots,
\rme^{2\pi\rmi k/\Nc})$ where $k=0,1,\dots\Nc-1$.  (See
figures~\ref{fig:pol}~(a) and (b) for the graphical representation of
the transformation.)  Because this ${\rm Z}_{\Nc}$ group is a centre
of the ${\rm SU}(\Nc)$ gauge group, the global symmetry under the
transformation~\eref{eq:center-trans} is called centre symmetry and
the Polyakov loop changes accordingly as $L \to z_k\, L$.  Hence, the
Polyakov loop expectation value is an order parameter for the
spontaneous breaking of centre symmetry. 

In general the expectation value of operator can be determined so
as to minimise the effective potential.  Therefore, the effective
action $\Gamma[\Phi]$ or the effective potential $V[\Phi]$ would
suffice to give the information on whether the system is in the
confined or deconfined phase.


\subsection{Perturbative approaches}

Because the perturbative calculations are valid at such high
temperature that gluons and quarks interact weakly, the deconfined
phase should be favoured in the perturbative regime and $\Phi\sim 1$
should be concluded.  This anticipation was first confirmed by in
\cite{Weiss:1980rj,Weiss:1981ev} in the SU(2) and SU(3) pure
Yang-Mills theories.  In the perturbative calculation it is more
convenient to formulate the effective potential not in terms of $\Phi$
directly but the phases of $L$ instead.

\begin{figure}
\begin{center}
 \includegraphics[width=0.4\textwidth]{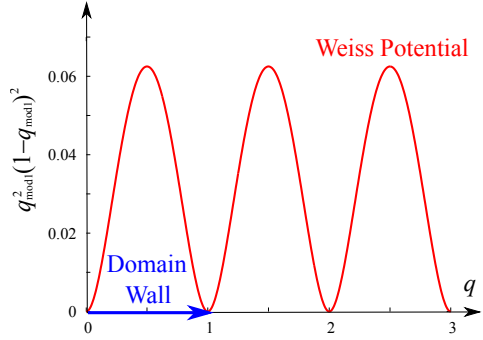}
\end{center}
\caption{SU(2) Weiss potential (up to the normalisation) as a function
  of $q$ (rescaled $A_4$, i.e.\ the phase of the Polyakov loop).  The
  domain wall configuration interpolates between two minima.}
\label{fig:domain}
\end{figure}

In this article we shall limit ourselves to the simple case of colour
SU(2) only.  The generalisation to colour SU(3) is straightforward.
Then, with an appropriate choice of the basis in colour space with
which $A_4$ is diagonal;
\begin{equation}
 A_4 = \frac{2\pi T}{g}\,q\, \Biggl(\begin{array}{cc}
  1 & 0 \\ 0 & -1 \end{array}\Biggr) \;.
\end{equation}
The Polyakov loop matrix and its trace can be expressed in terms of
$q$ as
\begin{equation}
 L = \Biggl(\begin{array}{cc}
  \rme^{\rmi \pi q} &  0 \\
  0 & \rme^{-\rmi \pi q}
 \end{array}\Biggr) \;,\qquad
 \Phi = \langle \cos(\pi q)\rangle
\label{eq:Phi}
\end{equation}
in the SU(2) case.  Then, performing the one-loop integration on top
of the $A_4$ background, one can find the following expression;
\begin{equation}
 V_{\rm eff}^{(1)}[q] = 2VT \int\frac{\rmd^3 p}{(2\pi)^3} \Bigl\{
  \beta p + \ln\bigl[ 1-\rme^{-\beta p + 2\rmi\pi q} \bigr]
  +\ln\bigl[ 1-\rme^{-\beta p - 2\rmi\pi q} \bigr] \Bigr\} \;,
\label{eq:Weiss-pre}
\end{equation}
where the Polyakov loop enters as an imaginary colour chemical
potential.  After dropping the zero-point energy and carrying the
momentum integration out, one can arrive finally at the perturbative
effective potential (namely the Weiss potential),
\begin{equation}
 V_{\rm eff}^{(1)}[q] = \frac{4\pi^2 V}{3\beta^4} \,
  \qmod^2 (1-\qmod)^2 \;.
\label{eq:Weiss}
\end{equation}
It is obvious from equation~\eref{eq:Weiss-pre} that
$V_{\rm eff}^{(1)}[q]$ is a periodic function of $q$ with the period
$1$, that is the reason why $\qmod$ appears in
equation~\eref{eq:Weiss}.  We remark that the Weiss potential is
conveniently expressed by means of the Bernoulli polynomials.
\Fref{fig:domain} is a sketch of this periodic potential with the
horizontal axis of $q$.  In view of this periodic structure, one
may well think of a tunnelling process that interpolates two minima.  We
will come to this point soon later.

The two-loop calculations were first attempted in references
\cite{Belyaev:1989yt,Enqvist:1990ae}, which turned out incomplete
because the one-loop correction to the Polyakov loop had been missing;
in the computation of the Polyakov loop, the tree-level relation,
$\langle\cos(\pi q)\rangle\approx \cos(\pi\langle q\rangle)$, is
insufficient for the two-loop potential.  In this way, a simple
prescription to treat the $A_4$-background as if it were the Polyakov
loop itself has a potential risk of pitfall in the non-perturbative
regime.  This correction was taken into the SU(2) calculation first
\cite{Belyaev:1991gh} and the SU($\Nc$) calculation later
\cite{KorthalsAltes:1993ca}.  Although the complete expression for the
SU($\Nc$) case has a complicated combination of the Bernoulli
polynomials, it is reduced to a simple form in the SU(2) case.  The
effective potential of the two-loop order is composed from two pieces;
$V_{\rm f}^{(2)}[q]$ from the direct contribution of the two-loop
integration and $V_{\rm p}^{(2)}[q]$ from the Polyakov-loop
renormalisation at one-loop order;
\begin{equation}
 V_{\rm eff}^{(2)}[q] = V_{\rm f}^{(2)}[q] + V_{\rm p}^{(2)}[q] \;.
\end{equation}
They are individually calculated and the SU(2) results read;
\begin{eqnarray}
 V_{\rm f}^{(2)}[q] = \frac{2\pi\as V}{\beta^4}\Biggl[
  \qmod^2(1-\qmod)^2 - \frac{2}{3}\qmod(1-\qmod) \Biggr] \;,\\
 V_{\rm p}^{(2)}[q] = -\frac{16\pi\as V}{3\beta^4}\Biggl[
  \qmod^2(1-\qmod)^2 - \frac{1}{4}\qmod(1-\qmod) \Biggr] \;.
\end{eqnarray}
Adding these two potentials up, we find that the latter terms inside
of the parentheses cancel out and the final expression simplifies as
\begin{equation}
 V_{\mathrm{eff}}^{(2)}[q] = -\frac{10\pi\as V}{3\beta^4}
  \qmod^2(1-\qmod)^2 \;.
\label{eq:Phi2}
\end{equation}
Therefore, interestingly enough, the correction of the two-loop order
modifies only the overall coefficient of the Weiss potential and does
not alter the functional form of the potential.  We can confirm that
the perturbative vacuum at $q=0$ is indeed a minimum of the potential
and then $\Phi=+1$ is concluded from equation~\eref{eq:Phi}, which is
not ruined by the correction~\eref{eq:Phi2}.

From the Weiss potential, in the case when quarks are absent, the
stable vacuum is degenerate at $q=n$ with $n$ being an integer,
reflecting centre symmetry.  These minima are, in fact, to be
connected by the centre transformation.  For general $q=n$ the
Polyakov loop takes a value of $(-1)^n$.  This degeneracy between
$\Phi=+1$ and $\Phi=-1$ would be broken by the presence of quarks.

The quark contribution to the Polyakov loop potential is known up to
the two-loop order \cite{KorthalsAltes:1993ca}.  We explain only the
one-loop result here.  The one-loop integration reads, apart from the
zero-point energy,
\begin{equation}
 V_{\rm quark}^{(1)}[q] = -4\Nf VT \int\frac{\rmd^3 p}{(2\pi)^3} \Bigl\{
  \ln\bigl[ 1+\rme^{-\beta p + \rmi\pi q} \bigr]
  +\ln\bigl[ 1+\rme^{-\beta p - \rmi\pi q} \bigr] \Bigr\} \;,
\label{eq:Weiss-quark-pre}
\end{equation}
for massless $\Nf$ flavours, which eventually amounts to
\begin{equation}
 V_{\mathrm{quark}}^{(1)}[q] = -\frac{8\Nf \pi^2 V}{3\beta^4}
  \biggl(\frac{q}{2}+\frac{1}{2}\biggr)_{\rm mod1}^2
  \Biggl[1-\biggl(\frac{q}{2}+\frac{1}{2}\biggr)_{\rm mod1}
  \Biggr]^2 \;.
\label{eq:Weiss-quark}
\end{equation}
It is worth mentioning that one can recover the above functional form
immediately by replacing $q\to q/2 + 1/2$ in the Weiss
potential~\eref{eq:Weiss}.  In this replacement an additional term
$1/2$ comes from the quantum statistics (boson or fermion) with which
the exponential term changes the sign.  Also, the argument is $q/2$
instead of $q$ because quarks belong to the colour fundamental
representation, while gluons are in the adjoint representation.  We
will take a closer look at quark effects when we discuss the model
studies at finite density in \sref{sec:models}.

Since the period is doubled as compared to the pure gluonic case, the
potential has no degeneracy between $\Phi=\pm 1$.  Thus, $\Phi=1$ (or
$q=0$) is more favoured than $\Phi=-1$, which is a consequence of the
explicit breaking of centre symmetry caused by quarks.  In the
presence of dynamical quarks, therefore, the state at $\Phi=-1$ is a
metastable vacuum.  It is shown \cite{Belyaev:1991cw}, however, that
the metastable state has physically unacceptable properties in
thermodynamics.  Different Polyakov-loop domains, whose interface is
the ${\rm Z}_{\Nc}$ domain wall
\cite{Bhattacharya:1990hk,Bhattacharya:1992qb}, are meaningful only in
Euclidean space-time (see \cite{Smilga:1993vb,Smilga:1994mf} for
detailed arguments against the physical interpretation of the
${\rm Z}_{\Nc}$ domain wall).  For the physical interpretation of the
Polyakov loop in Minkowskian space-time the 't~Hooft loop as we will
discuss below is of special importance \cite{KorthalsAltes:1999xb}.

\begin{figure}
\begin{center}
 \includegraphics[width=0.4\textwidth]{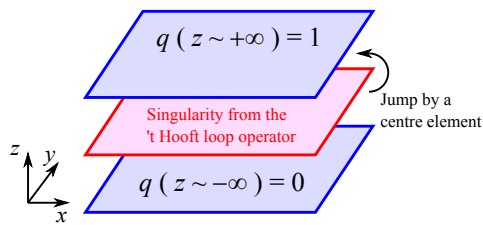}
\end{center}
\caption{Schematic figure of the 't~Hooft loop.  A singularity on the
  surface $\Sigma$ twists the boundary condition for the Polyakov loop
  by a centre element.}
\label{fig:thooft}
\end{figure}

So far, we have considered only homogeneous configuration of the
Polyakov loop background.  One interesting application of the Weiss
potential is the formation of the domain wall that is an inhomogeneous
object in space.  It is then convenient to introduce what is called
the 't~Hooft loop \cite{'tHooft:1977hy} (which measures the
chromo-electric flux) besides the Wilson loop (which measures the
chromo-magnetic flux).  We denote the 't~Hooft loop along the contour
$C$ as $V(C)$ and then it should satisfy the operator relation;
\begin{equation}
 V^\dagger(C) W(C') V(C) = \rme^{2\rmi \pi Lk(C,C')/\Nc} W(C')
\label{eq:VW}
\end{equation}
with a centre-element coefficient in the right-hand side and Gauss'
link number $Lk(C,C')$ of the two contours $C$ and $C'$.
Mathematically $Lk(C,C')$ can be expressed as \cite{Reinhardt:2002mb}
\begin{equation}
 Lk(C,C') = \frac{1}{4\pi}\oint_{C}\rmd x_i \oint_{C'}\rmd y_j
 \,\epsilon_{ijk}\,\frac{x_k-y_k}{|\bi{x}-\bi{y}|^3} \;,
\end{equation}
which shall be easily understood from Amp\`{e}re's law with the
magnetic field given by Biot-Savart's law in electromagnetism.
From the fact that the 't~Hooft loop counts the chromo-electric flux,
with a certain choice of the colour direction (which can be taken as
the 3-rd direction in the SU(2) case without loss of generality), the
explicit form of the 't~Hooft loop could be given as
\cite{KorthalsAltes:1999xb}
\begin{equation}
 V(C) = \exp\Biggl[ \frac{2\pi \rmi}{g}\int_{\Sigma} \rmd^2 S^i
 E_i^3 \Biggr] \;,
\end{equation}
where $\Sigma$ is the two-dimensional sheet enclosed by $C$, i.e.\
$\partial\Sigma=C$.  In the computation of the 't~Hooft loop
expectation value, the insertion of this operator to the functional
integration induces a delta-function singularity (Dirac surface) on
$\Sigma$, which makes a twist on the boundary condition for the
Polyakov loop by a centre element $z_k$ (that is $-1$ for the SU(2)
group).  We illustrate a schematic picture in \fref{fig:thooft}.

With the choice of $C$ in the whole $x$-$y$ plane, the surface
$\Sigma$ spans over the $x$-$y$ plane at a certain ($z_0,\,t_0$).  The
expectation value of the 't~Hooft loop is then expressed as
\cite{deForcrand:2005pb}
\begin{equation}
 \langle V(C)\rangle = \frac{Z_{\rm tbc}}{Z_{\rm pbc}} \;,
\label{eq:thooft_vev}
\end{equation}
where $Z_{\rm pbc}$ is the partition function with the periodic
boundary condition in the $z$-direction, while $Z_{\rm tbc}$ has a
twisted boundary condition for the Polyakov loop.

Hence, to compute the expectation value of the 't~Hooft loop, an
effective action in terms of $q$ is necessary including the derivative
terms, that is given by
\begin{equation}
 \Gamma_{\mathrm{eff}}[q] = L_x L_y \int_0^{L_z}\rmd z \Biggl\{
  \frac{2\pi^2}{g^2\beta}\Biggl[ \frac{\rmd q(z)}{\rmd z} \Biggr]^2
  +\frac{4\pi^2}{3\beta^3}
  \,q(z)^2 \, \bigl[ 1 - q(z) \bigr]^2 \Biggr\}\;,
\end{equation}
at one-loop order.  This potential term is nothing but the Weiss
potential and the derivative term is from the tree-level action.  We
note that the quantum corrections to the derivative term have been
evaluated beyond the derivative expansion and a possibility of
spatially inhomogeneous configuration of the Polyakov loop has been
suggested \cite{Fukushima:2000ww}.

In the presence of the 't~Hooft loop, the twisted boundary condition
is that $q\to 0$ at $z\to -\infty$ and $q\to 1$ at $z\to +\infty$.
The 't~Hooft loop is a creation operator of centre-domain interface,
therefore.  The classical solution associated with the above effective
action is
\begin{equation}
 q_{\rm c}(z) = \frac{1}{1+\exp[-\sqrt{2/3}gTz]} \;,
\end{equation}
which satisfies the boundary condition; $q(z\to-\infty)=0$ and
$q(z\to+\infty)=1$.  Then, the effective action takes a finite value
that is \cite{Bhattacharya:1990hk},
\begin{equation}
 \Gamma_{\mathrm{eff}}[q_{\rm c}(z)]
  = \frac{4\pi^2}{3\sqrt{6} g \beta^2} L_x L_y
  = \sigma_{\rm t} L_x L_y \;,
\end{equation}
From this with equation~\eref{eq:thooft_vev}, the expectation value of
the 't~Hooft loop at the one-loop order is,
\begin{equation}
 \langle V(C)\rangle = \exp[-\sigma_{\rm t} \Sigma(C)]
\end{equation}
with $\Sigma(C)=L_x L_y$, which shows the area law in the
deconfinement phase.  We see that the 't~Hooft loop has behaviour
opposite to the Wilson loop, and the 't~Hooft loop plays the role of
disorder parameter.  Together with $W(C)$ and $V(C')$ the state of
matter is characterised in more details as
\begin{equation}
 \mbox{\underline{Confined Phase}}
   \qquad \langle W(C)\rangle \sim \rme^{-\sigma_{\rm w} \Sigma(C)} \;,
   \quad \langle V(C')\rangle \sim \rme^{-\sigma_{\rm t}' L(C')} \;.
\end{equation}
In the deconfinement phase or the Higgs phase, in contrast, the
behaviour is opposite;
\begin{equation}
 \mbox{\underline{Higgs Phase}}
   \qquad \langle W(C)\rangle \sim \rme^{-\sigma_{\rm w}' L(C)} \;,
   \quad \langle V(C')\rangle \sim \rme^{-\sigma_{\rm t} \Sigma(C')} \;.
\end{equation}
In the Higgs phase at high temperature, there is no massless particle.
It should be noted that all gluons are massive due to the thermal
screening mass.  Here, we can also consider the third possibility;
\begin{equation}
 \mbox{\underline{Partial Higgs Phase}}
 \qquad\!\! \langle W(C)\rangle \sim \rme^{-\sigma_{\rm w} \Sigma(C)} \;,
 \quad\!\! \langle V(C)\rangle \sim \rme^{-\sigma_{\rm t} \Sigma(C)} \;,
\end{equation}
which represents the partial Higgs phase and confinement still
remains.  The last possibility that both the Wilson and the 't~Hooft
loops show the perimeter law is excluded from the operator algebra
\eref{eq:VW} \cite{'tHooft:1977hy}.

Here we may understand that it is the spatial 't~Hooft loop that
exists in Minkowskian space-time and the real-time counterpart of the
${\rm Z}_{\Nc}$ interface is something that is created by the
't~Hooft loop.  The Polyakov loop in the real-time dynamics should be
understood along this line \cite{Hidaka:2009hs}.

We have, so far, discussed the behaviour of the Wilson loop or the
Polyakov loop as the order parameter and the spatial 't~Hooft loop as
the disorder parameter.  The spatial Wilson loop is also an
interesting quantity.  For completeness we shall give a brief
description about the spatial Wilson loop.  It always shows the area
law regardless of the temperature.  This can be understood in the
3-dimensional effective theory of QCD at high temperature
\cite{Braaten:1994na} as a result of the dimensional reduction
\cite{Appelquist:1981vg}.

Integrating all ``hard'' modes out with non-zero Matsubara frequency
at high temperature leaves a 3-dimensional effective theory of the
``soft'' length scales $> (gT)^{-1}$.  This effective theory is
commonly referred to as Electrostatic QCD (EQCD), that is defined by
the Lagrangian,
\begin{equation}
\hspace{-4em}
 S_{\rm EQCD} = \int \rmd^3 x\,\Biggl\{ \frac{1}{2}\tr
  F_{ij} F_{ij} + \tr\bigl( D_i A_0 \bigr)^2
  +m_{\rm E}^2 \tr A_0^2 + \lambda_{\rm E}\bigl[ \tr(A_0^2)^2 \bigr]
  +\bar{\lambda}_{\rm E}\tr A_0^4 \Biggr\} \;.
\end{equation}
The electrostatic field $A_0(\bx)$ is static and adjoint scalar in
colour space.  The matching parameters in EQCD are calculated
\cite{Braaten:1995jr} and the three-loop level is still in progress
\cite{Laine:2005ai,Moller:2010xw}.  The leading-order results are;
\begin{equation}
\hspace{-2em}
 g_{\rm E}^2 = g^2(T) T \;,\quad\!
 m_{\rm E}^2 = \frac{\Nc}{3}g^2(T) T^2 \;,\quad\!
 \lambda_{\rm E}=\frac{g^4(T)T}{4\pi^2} \;,\quad\!
 \bar{\lambda}_{\rm E}=\frac{\Nc g^4(T)T}{12\pi^2} \;.
\end{equation}
Because $A_0(\bx)$ is a heavily massive mode at high temperature,
integrating $A_0$ out leads to an effective theory of the
``ultrasoft'' length scales $>(g^2 T)^{-1}$, which is called
Magnetostatic QCD (MQCD).  This MQCD is defined by the Lagrangian,
\begin{equation}
 S_{\rm MQCD} = \int\rmd^3 x\, \frac{1}{2}\tr F_{ij} F_{ij} \;.
\end{equation}
This is a confining theory with the magnetic coupling constant, which
is
\begin{equation}
 g_{\rm M}^2 = g_{\rm E}^2 = g^2(T) T \;,
\end{equation}
at the leading order.  From the dimensional reason the string tension
associated with this 3-dimensional effective theory (namely, the
string tension measured by the spatial Wilson loop) is to be
parametrised as
\begin{equation}
 \sigma_{\rm s} = c^2 g_{\rm M}^4 \;.
\end{equation}
The determination of $c$ requires full non-perturbative evaluation,
and the Monte-Carlo simulation of the pure gluonic theory results in
$c\approx 0.553(1)$ \cite{Karsch:1994af,Teper:1998te}, which is also
confirmed by later simulation \cite{Lucini:2002wg}.

Although systematic resummation programs in EQCD and MQCD are on the
track, the Polyakov loop effects and confinement physics are not
incorporated in a satisfactory manner (see reference
\cite{Hidaka:2009hs} for some attempts and also \cite{Vuorinen:2006nz}
for another idea).  It is quite difficult to investigate the nature of
deconfinement phase transition in the perturbative approaches.


\subsection{Non-perturbative methods at work}

Theoretical researches on the confining properties near and below
$\Tc$ require non-perturbative extensions of the method.  The
lattice-QCD simulations are the most successful as long as the quark
chemical potential is sufficiently smaller than the temperature.  For
recent developments in the lattice-QCD calculations there are a number
of nice reviews (see reference \cite{DeTar:2009ef} for example).  In
this review article we shall focus on some of analytical approaches.


\subsubsection{Strong-coupling expansion:}

The deconfinement phase transition can be formulated
non-perturbatively in the limit of the strong coupling constant,
$g^{-1}\to 0$, which was first elucidated in the Hamiltonian formalism
in reference \cite{Polyakov:1978vu}.  The same conclusion is readily
obtained in the formalism of functional integration
\cite{Polonyi:1982wz,Gross:1983ju}.

\begin{figure}
\begin{center}
 \includegraphics[width=0.35\textwidth]{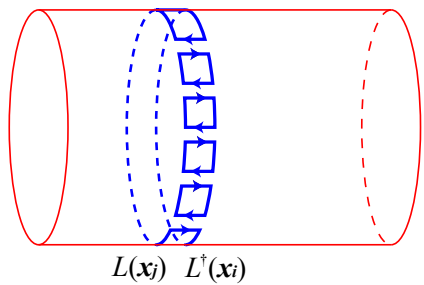}
\end{center}
\caption{Leading order contribution to the Polyakov loop effective
  action in the strong-coupling limit.}
\label{fig:finite-exp}
\end{figure}

In the leading order of the plaquette expansion as sketched in
\fref{fig:finite-exp}, the effective action in terms of the Polyakov
loop reads,
\begin{equation}
 S_{\rm pol}[L] = -\rme^{-\sigma a/T}
  \sum_{\rm n.n} \tr L^\dagger(\bi{x}_i) \tr L(\bi{x}_j) \;,
\label{eq:pol-matrix}
\end{equation}
which describes a hopping interaction between adjacent Polyakov
loops.  Here $a$ is the lattice spacing.  This action actually defines
a spin-like theory of the Polyakov loop matrix;
\begin{equation}
 Z = \int \mathcal{D}L\, \rme^{-S_{\rm pol}[L]} \;.
\end{equation}
Here $\mathcal{D}L$ represents the functional integral with the group
invariant (Haar) measure.  The theoretical content of this matrix
model itself is very intriguing \cite{Kogut:1981ez}.  In the same
manner as the mean-field treatment (or the so-called molecular-field
approximation) of spin systems, it is possible to formulate the
spontaneous breaking of centre symmetry and $\Phi$ takes a finite
value when the spin interaction becomes large at sufficiently high $T$
\cite{Ilgenfritz:1984ff,Gocksch:1984yk,Fukushima:2002ew,Fukushima:2003fm}.

Although it is much simpler than the molecular-field approximation,
the tree-level potential is already useful to describe the
deconfinement phase transition.  In this prescription the traced
Polyakov loop in the action~\eref{eq:pol-matrix} is simply replaced
by the expectation value $\Phi$, and an additional contribution comes
from the Haar measure in the functional integration, i.e.
\begin{equation}
 V_{\mathrm{eff}}[\Phi] = -6V\Nc^2\, \rme^{-\sigma a/T} \bar{\Phi}\Phi
  -\ln\mathcal{M}_{\mathrm{Haar}}[\Phi] \;,
\end{equation}
where the Haar measure for the SU($\Nc$) group is given by
\begin{equation}
\hspace{-2em}
 \ln \mathcal{M}_{\mathrm{Haar}} = \cases{
  V\,\ln\bigl[ 1 - \bar{\Phi}\Phi \bigr] \;,
   \qquad\qquad\qquad\qquad\qquad\quad\: \mbox{(for $\Nc=2$)} \\
  V\,\ln\bigl[ 1-6\bar{\Phi}\Phi+4(\bar{\Phi}^3 + \Phi^3)
   -3(\bar{\Phi}\Phi)^2 \bigr] \;.
  \quad\, \mbox{(for $\Nc=3$)} }
\end{equation}
It is important to note that these Haar measures favour the confining
state at $\Phi=0$.  Moreover, the perturbative vacuum at $\Phi=\pm1$
has an infinitely high barrier, which is cancelled by the longitudinal
gluon loop in the perturbative calculation.  Thus, the Haar measure
could play an essential role in the realisation of confinement
\cite{Gocksch:1993iy,Lenz:1998qk}.  Together with this Haar measure
contribution and the spin interaction term, a phase transition takes
place on the mean-field level and it is of second order for $\Nc=2$
and of first order for $\Nc=3$.  Here, we distinguish the
anti-Polyakov loop,
$\bar{\Phi}=\langle\tr L^\dagger\rangle/\Nc$, from $\Phi$;  they are
just identical at zero baryon density but a discrepancy between them
arises from finite-density effects and has much to do with the sign
problem.  We shall return to this problem in the next section.

The history of the investigations on chiral symmetry restoration in
the strong-coupling expansion is as long as that of deconfinement
physics, which is summarised in a review \cite{Fukushima:2003vi}.
This section is devoted mainly to deconfinement physics and we will
later look over the physics implications of chiral dynamics in
\sref{sec:density}.

\begin{figure}
\begin{center}
 \includegraphics[width=0.47\textwidth]{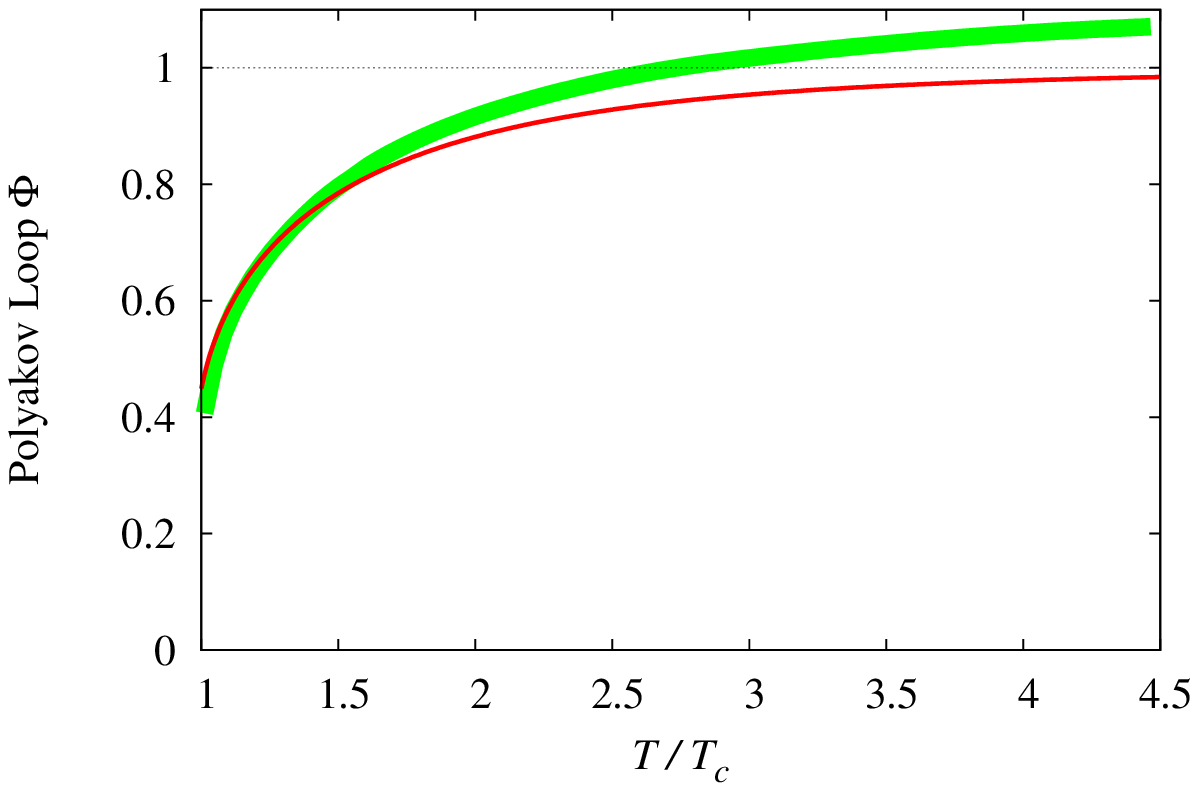}\hspace{0.5em}
 \includegraphics[width=0.47\textwidth]{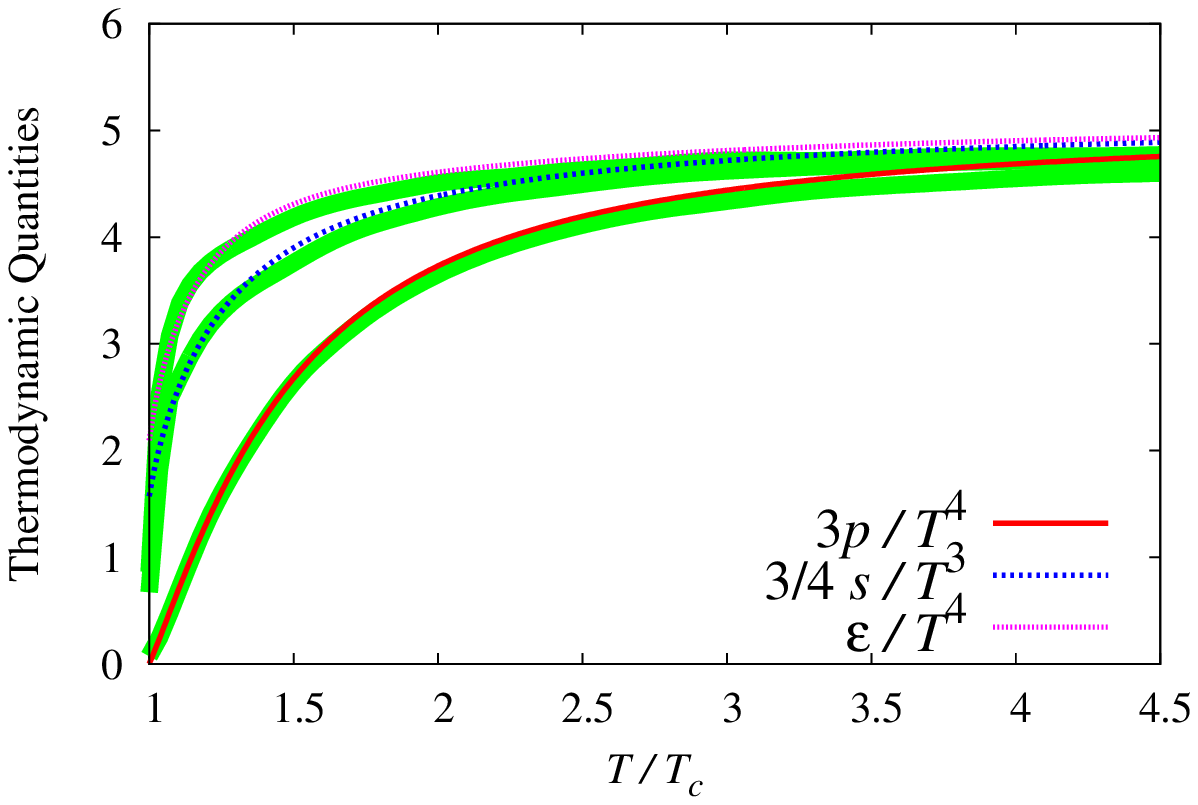}
\end{center}
\caption{Left: Bands represent the lattice data for the Polyakov loop
  and the solid curve is the fitted result from
  equation~\eref{eq:pol-para}.\ \ Right: Pressure, the entropy density
  and the internal energy density from lattice data and the fitted
  result.  The lattice data is taken from reference
  \cite{Gupta:2007ax}.  Similar figures are found in
  \cite{Fukushima:2010zz}.}
\label{fig:pure}
\end{figure}

Inspired by the functional form from the strong-coupling analysis, one
can adopt the following Ansatz to fit the pressure in the pure gluonic
sector;
\begin{equation}
 V(\Phi) = -\frac{a(T)}{2}\bar{\Phi}\Phi + b(T)\ln\Bigl[ 1
  -6\bar{\Phi}\Phi + 4(\bar{\Phi}^3+\Phi^3) - 3(\bar{\Phi}\Phi)^2
  \Bigr] \;,
\label{eq:Pol-pot}
\end{equation}
where a set of parameters, $a(T)/T^4 = 3.51-2.47t^{-1}+15.2t^{-2}$ and
$b(T)/T^4 = -1.75t^{-3}$ with $t=T/T_{\rm c}$, can reproduce the
lattice data well \cite{Roessner:2006xn}.  In this parametrisation
there are only three free variables because one of four is constrained
by the Stefan-Boltzmann law.  It is amazing that not only the Polyakov
loop but also the pressure estimated from
\begin{equation}
 \frac{\rmd V(\Phi)}{\rmd \Phi}\Biggr|_{\Phi=\Phi_0} = 0 \;,\qquad
 P = V(\Phi=\Phi_0) \;,
\label{eq:pol-para}
\end{equation}
simultaneously agree well with the lattice data \cite{Gupta:2007ax},
as seen in the plots in \fref{fig:pure}.

The agreement is impressive for only three fitting parameters, except
for the Polyakov loop at very high temperature.   The Polyakov loop
from the lattice data exceeds unity there, which is caused by the
renormalisation effect on the Polyakov loop
\cite{Polyakov:1980ca,Creutz:1980hb,Kaczmarek:2002mc,Gupta:2007ax}.
It is known that the UV divergence in the Polyakov loop is absorbed by
the charge renormalisation as
\begin{equation}
 \Phi^{\rm ren}(T) = \bigl[ Z(g^2) \bigr]^{\Nt}
  \Phi^{\rm bare}(g^2,\Nt) \;.
\end{equation}
The renormalisation constant is fixed at a reference temperature, that
corresponds to the renormalisation condition.  Once this is done in a
certain scheme, a renormalisation constant at a different temperature
is fixed by a different $\Nt$, which in turn leads to the
renormalisation constant at a different coupling $g$ (or lattice
spacing).  In this way the renormalised Polyakov loop is calculated for
all temperatures through the iterative procedure.  It is a non-trivial
question how to incorporate the renormalisation effect in the Polyakov
loop model \cite{Dumitru:2003hp}.


\subsubsection{Large-$\Nc$ QCD:}

The confinement-deconfinement transition is well-defined only in the
pure Yang-Mills theories without quarks in the colour fundamental
representation or in the limit of infinite quark mass.  Otherwise, in
the presence of dynamical quarks, centre symmetry is explicitly broken
and the Polyakov loop always takes a finite value; $\Phi\neq0$.  The
Polyakov loop correlation function, in other words, does not decay
exponentially at large distances due to pair creation of quark and
anti-quark.

Even in the presence of dynamical quarks, however, there is another
limit in which the deconfinement transition is well-defined.  That is,
increasing the number of gluons instead of decreasing the number of
quarks.  In fact the quark contribution is more suppressed than
gluons with large number of colours $\Nc\to\infty$
\cite{'tHooft:1973jz,Witten:1979kh}, and eventually, in the limit of
infinite colours a smooth crossover of deconfinement turns into a
sharp phase transition.

If we see the pressure of finite-$T$ hadronic matter, on the one hand,
it is of $O(1)$ in the $\Nc$ counting.  There are gluons of
$O(\Nc^2)$, on the other hand, and the pressure of deconfined matter
is of $O(\Nc^2)$.  This is dominant over the quark contribution of
$O(\Nc)$.  Therefore, in the limit of $\Nc\to\infty$, the quark
contribution becomes negligible and the quenched approximation works
correctly.  The pressure then jumps from $O(1)$ to $O(\Nc^2)$ when
the system goes through the phase transition from the hadronic to the
deconfined phases.  This means that the location of the phase
transition has no ambiguity unlike the pseudo-critical temperature of
crossover at $\Nc=3$.

It is a subtle question what the order of the phase transition would
be in the $\Nc\to\infty$ limit.  One might have thought that the phase
transition should be of first order simply because of a big jump in
the pressure.  It is, however, possible to have a second (or higher)
order phase transition with a pressure that is continuous but
increasing rapidly.  Thus we can naively think of two possibilities:  (1)
The phase transition (or crossover) is continuous for any $\Nc$ and
the $\Nc\to\infty$ limit makes it of second order.  (2)  The phase
transition at $\Nc\to\infty$ is of first order and there is a critical
number of colours at which the phase transition is of exact second
order.

The fact seems to be more complicated.  In the large-$\Nc$ limit of
the Polyakov loop matrix model another possibility has been suggested;
the Gross-Witten point might be realised at $\Nc\to\infty$
\cite{Dumitru:2004gd}.  Then the effective potential is flat in the
region $0\le \Phi < 1/2$ and starts increasing for $\Phi\ge 1/2$.  The
Polyakov loop jumps from $0$ to $1/2$ at the critical point.  This is
an unconventional point because no interface tension is needed for a
jump between $\Phi=0$ and $\Phi=1/2$, which would turn into a
continuous transition immediately with an infinitesimal background.

The large-$\Nc$ approach is useful as long as dropping dynamical
quarks off is not critically harmful.  It is an interesting
theoretical challenge to apply the large-$\Nc$ argument for
finite-density problems.


\subsubsection{Holographic model:}

The application of the AdS/CFT correspondence has become an important
building block of hot and dense QCD physics.  The idea is that the
weak-coupling Type-IIB supergravity theory on
${\rm AdS}_5\times {\rm S}_5$ is equivalent to the strong-coupling
$\mathcal{N}=2$ super Yang-Mills theory on the boundary of
${\rm AdS}_5$ space.  Especially the presence of the QGP is translated
into a non-extremal black-hole solution (see references
\cite{Witten:1998zw,Klebanov:2000me} for reviews).  This technique is
quite useful to examine non-perturbative aspects of strong-coupling
gauge theories.  The problem in the application to QCD physics is that
QCD is neither conformal invariant nor supersymmetric.  There are a
number of theoretical attempts to design the black-hole solution so
that it can mimic QCD thermodynamics
\cite{Andreev:2007zv,Gursoy:2008bu,Gubser:2008ny} (see also an extensive
comparison with the lattice simulation \cite{Panero:2009tv}) with a hope
to establish the AdS/QCD model.

One of the simplest ways to introduce a mass scale is to use the
following Ansatz for the five-dimensional background geometry (in
Einstein frame);
\begin{equation}
 \rmd s^2 = \rme^{c z^2}\frac{L^2}{z^2} \Bigl[ -f(z)\, \rmd t^2
  + \rmd \vec{x}^2 + f^{-1}(z)\, \rmd z^2 \Bigr]
\label{eq:soft-wall}
\end{equation}
with $f(z)=1-(z/z_{\rm h})^4$, which describes a Schwarzschild-type
black hole along the fifth coordinate $z$.  The corresponding Hawking
temperature is
\begin{equation}
 T = \frac{1}{\pi z_{\rm h}} \;,
\end{equation}
that is interpreted as the QGP temperature in the gauge theory side.
The dilaton potential contains a dimensional parameter $c$ that plays
the role of QCD scale in this model.  Such a model is usually referred
to as the soft-wall model (usually defined in string frame) and is
quite successful to give a semi-quantitative description of the Regge
trajectory in the vector-meson channel \cite{Karch:2006pv} (see also
reference \cite{Ghoroku:2005vt} for a related approach).  This type of
approach is generally called the ``bottom-up'' model.

Interestingly enough, the deconfinement phase transition is clearly
identified in such a holographic setup.  The QGP at high temperature is
featured by the metric~\eref{eq:soft-wall}, whereas confined matter at
zero temperature is described by the metric with $z_{\rm h}\to\infty$
(and thus $T\to 0$).  Then, the five-dimensional actions associated
with respective metrics determine which state is energetically
favoured.  In the soft-wall model the critical temperature $\Tc$ has
been obtained as $\Tc\simeq \sqrt{c}/\pi$
\cite{Andreev:2006eh,Herzog:2006ra} where the Hawking-Page transition
takes place.  It is noteworthy that this phase transition of
deconfinement is almost always of first order (see
\cite{Gursoy:2008bu} for a possibility of exception).

The Polyakov loop expectation value is calculated by the string
world-sheet area that is minimised with the boundary along the
Polyakov loop \cite{Andreev:2009zk}.  It follows that
\begin{equation}
 \Phi(T) = \exp\Biggl[ a - b\Biggl\{ \sqrt{\pi}
  \frac{\Tc}{T} \mathrm{Erfi}\biggl(\frac{\Tc}{T}\biggr) + 1
  -\rme^{(\Tc/T)^2} \Biggr\} \Biggr] \;,
\end{equation}
where $a$ is a normalisation constant and $b=R^2/2\alpha'$ is a
parameter in the Nambu-Goto string action.  A choice of $a=0.10$ and
$b=0.72$ fits the lattice data very well.

As far as the Polyakov loop behaviour and bulk thermodynamic
quantities are concerned, the results from the holographic approach
are no better than the strong-coupling expansion as we have seen in
equation~\eref{eq:pol-para}.  It should be an advantage in the
holographic model that some quantities that cannot be calculated on
the lattice can be calculated easily such as the transport
coefficients \cite{Policastro:2001yc,Kovtun:2004de}.  Indeed, the
computation of the shear viscosity in the strong-coupling expansion
has not been successful particularly including the effect of the
deconfinement transition \cite{Jakovac:2008ft}.


\section{Baryon-Rich State of QCD}
\label{sec:density}

Historically speaking, the possibility of deconfined gluons and quarks
was pointed out first for physics not at high temperature but at high
density in the context of neutron star structure
\cite{Itoh:1970uw,Collins:1974ky}.  It is, however, a non-trivial
question whether the QCD running coupling constant really gets smaller
at higher baryon density, as compared to the finite-$T$ case in
\sref{sec:temperature}.  Because quark excitations are allowed only
outside of the Fermi sphere if $T$ is small, dynamical quarks must
carry as large momentum as the quark chemical potential $\muq$.
The important point is that the relevant scale in $\as(\mu)$ in
equation~\eref{eq:running} is not the momentum of quarks but that of
the exchanged gluon between quarks.  It is still possible for
fast-moving quarks to emit and absorb soft gluons, for which $\mu$
could be small and $\as(\mu)$ could be substantially large.  Such soft
processes are, however, screened immediately due to quark polarisation
effects that induce a screening mass $\sim g\muq$ on gluons.  In this
way, in effect, one can regard sufficiently high-density matter of QCD
as a weak-coupling system.  This way of understanding is challenged
recently by the large-$\Nc$ approach to the QCD phase diagram as we
will see later in \sref{sec:largeNc-density}.

It is an ongoing experimental project to explore the state of QCD
matter in a wide range of temperature and baryon density by varying
the collision energy $\sqrt{s_{_{NN}}}$.  From the phenomenological
analysis using the thermal Statistical Model
\cite{Cleymans:2005xv,Becattini:2005xt,Andronic:2008gu} a set of the
temperature $T$ and the baryon chemical potential $\muB$ ($=3\muq$) at
which the chemical composition of particle species is frozen has been
extracted.  These ``chemical freeze-out points'' are very well
parametrised by \cite{Cleymans:2005xv}
\begin{equation}
 \Tf(\muB) = a - b \muB^2 - c \muB^4
\label{eq:Tf}
\end{equation}
with $a=0.166\pm0.002\GeV$, $b=0.139\pm0.016\GeV^{-1}$ and
$c=0.053\pm0.021\GeV^{-3}$.  Also the baryon chemical potential at
chemical freeze-out is parametrised as a function of the collision
energy as
\begin{equation}
 \muf(\sqrt{s_{_{NN}}}) = \frac{d}{1+e \sqrt{s_{_{NN}}}}
\end{equation}
with $d=1.308\pm0.028\GeV$ and $e=0.273\pm0.008\GeV^{-1}$, from which
one can convert equation~\eref{eq:Tf} into $\Tf(\sqrt{s_{_{NN}}})$
easily.

\begin{figure}
\begin{center}
\includegraphics[width=0.4\textwidth]{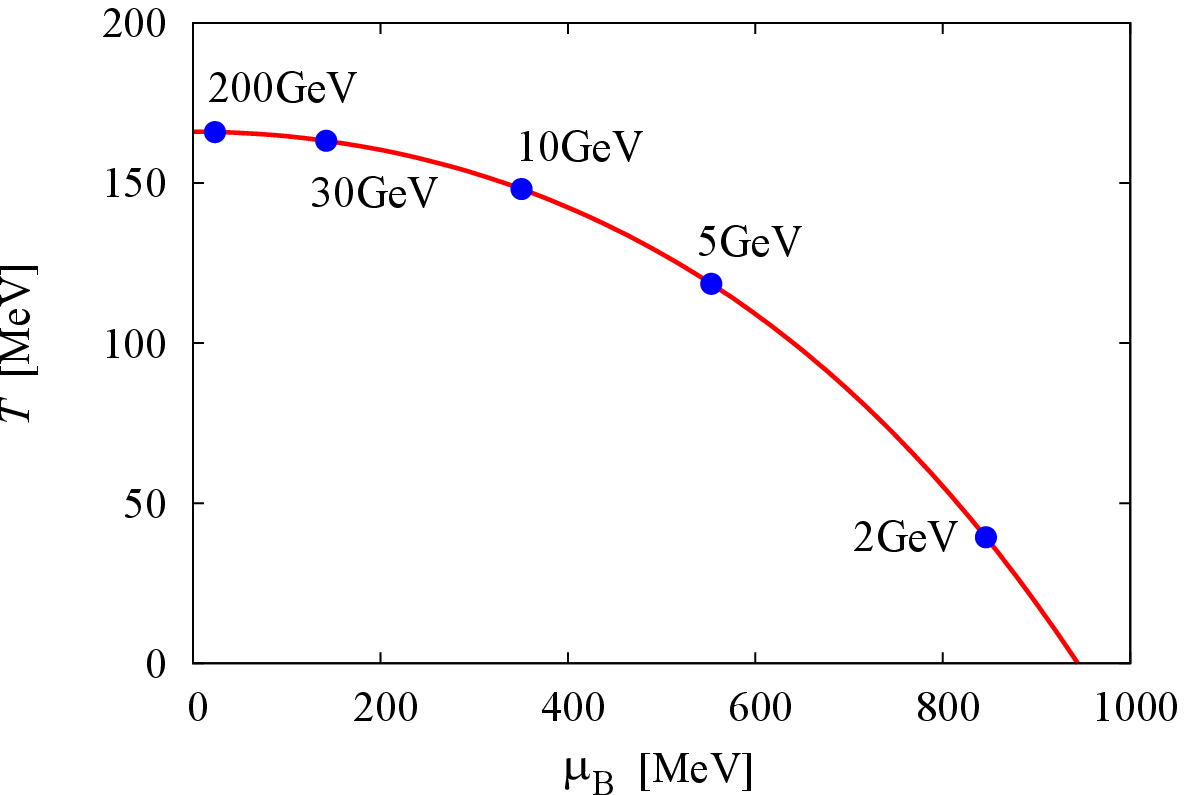} \hspace{1em}
\includegraphics[width=0.53\textwidth]{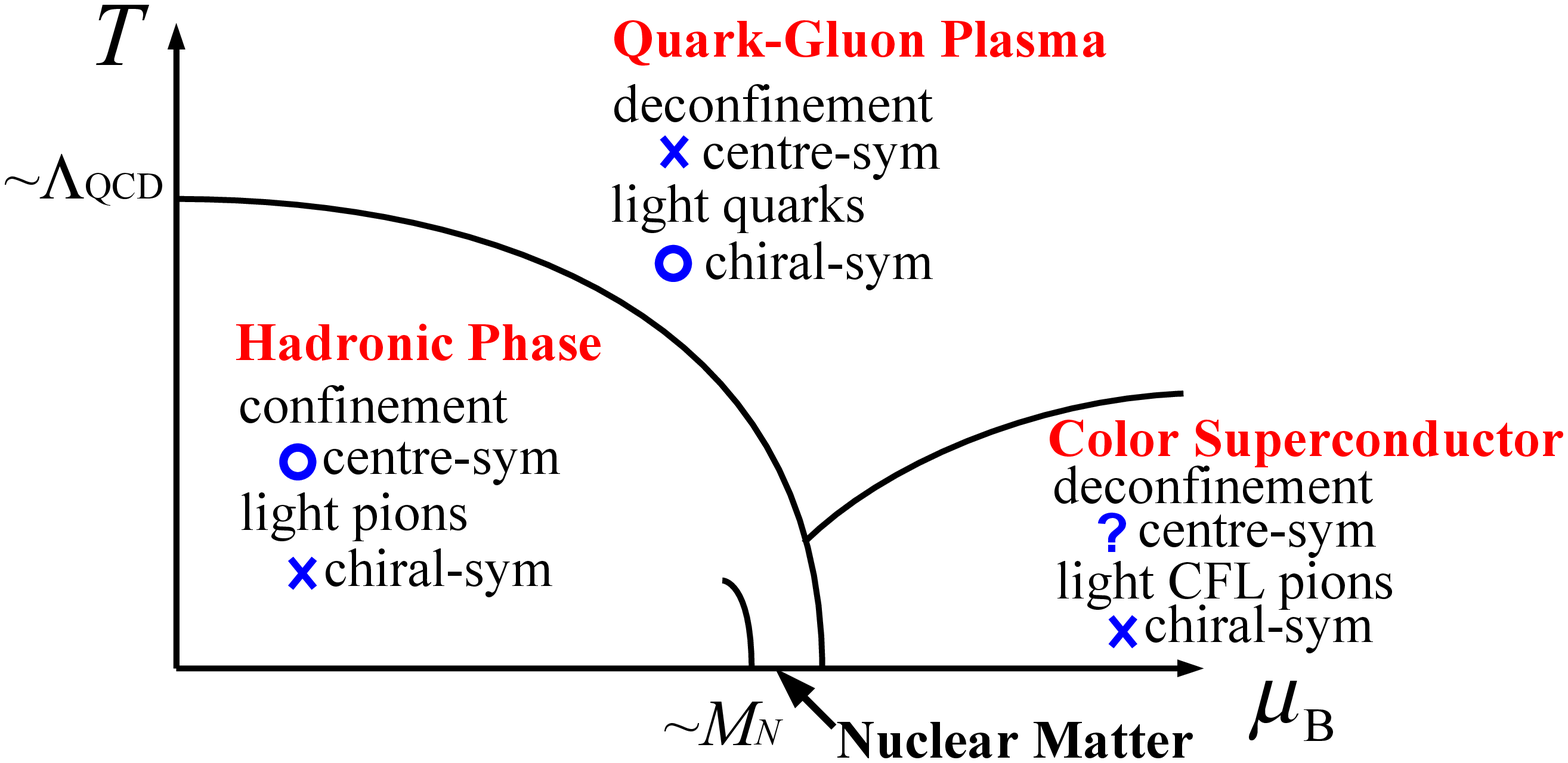}
\end{center}
\caption{Left: Chemical freeze-out line.  In the figure arbitrary five
  points are picked up to show the collision energy dependence of the
  chemical freeze-out point.\ \ Right: Minimal structure of the QCD
  phases in the $\muB$-$T$ plane.  The critical $T$ is of order
  $\LQCD$ and the critical $\muB$ is near the nucleon mass which is of
  order $\Nc\LQCD$.}
\label{fig:minimal}
\end{figure}

These parametrisations are useful to relate theoretical predictions
(for example, fluctuations of the charged particles, the baryon number
etc; see reference \cite{Karsch:2010ck}) to experimental observable as
a function of not only $\muB$ but also $\sqrt{s_{_{NN}}}$ that is
under experimental control.

It is a natural anticipation that this chemical freeze-out line, which
is plotted in the left of \fref{fig:minimal}, is related to the colour
deconfinement phenomenon that causes a rapid increase in the particle
number density, and thus the multiparticle scattering rate
\cite{BraunMunzinger:2003zz}.  We note that chiral symmetry
restoration is not taken into consideration at all in the assumption 
of the thermal Statistical Model.  In this way a part of the
theoretically conjectured QCD phase diagram (the right of
\fref{fig:minimal}) has been supported by experimental data, which
should be complemented further by a line associated with colour
superconductivity (see reference \cite{Alford:2007xm} for a modern
review and references therein).  We should emphasise that
\fref{fig:minimal} shows only the minimal structure of possible QCD
phases.  There are many other possibilities on top of this minimal
topology, some of which are summarised in my previous review
\cite{Fukushima:2010bq}.

Progresses in the experimental activities towards QCD matter at higher
baryon density are strongly needed by theorists.  Unlike the case at
high $T$ and small $\muB$, theoretical works have not been successful
in making any robust prediction on the baryon-rich state of QCD matter.
It is absolutely necessary to constrain proposed theoretical
possibilities from the experimental point of view.


\subsection{Perturbative approaches and problems}

Let us first consider an extension of the Weiss
potential~\eref{eq:Weiss} to the finite-density case.  We can perform
the perturbative integration as in the standard procedure with a
quark chemical potential $\muq$ introduced
\cite{Freedman:1976xs,Freedman:1976dm,Freedman:1976ub}.  It seems at a
first glance that such extension is straightforward, but it is not so
simple once the Polyakov loop background is involved.

The loop corrections to the Weiss potential have been evaluated with
dynamical (massless) quarks.  The one-loop contribution from massless
quarks is modified by $\muq$ from equation~\eref{eq:Weiss-quark-pre}
into
\begin{equation}
 \hspace{-5em}
 V_{\rm quark}^{(1)}[q] = -2\Nf VT \int\frac{\rmd^3 p}{(2\pi)^3} \sum_j
  \Bigl\{ \ln\bigl[ 1+\rme^{-\beta p +\beta\muq + \rmi\pi q_j} \bigr]
  +\ln\bigl[ 1+\rme^{-\beta p -\beta\muq - \rmi\pi q_j} \bigr]
  \Bigr\} \;,
\label{eq:Weiss-density-pre}
\end{equation}
which is generalised from SU(2) to SU($\Nc$) with $j$ running
from $1$ to $\Nc$ (and $\sum_j q_j=0$).  The first and second
logarithms represent the particle and the anti-particle excitations,
respectively.  Then, the result after the momentum integration is
obtained immediately \cite{KorthalsAltes:1999cp} by the replacement of
\begin{equation}
 q_j \;\to\; q_j - \rmi \frac{\muq}{\pi T}
\end{equation}
in equation~\eref{eq:Weiss-quark}.

This final result is simple but astonishing.  Unless the gauge group
is SU(2), the effective potential generally takes a complex value.
How can one determine the energetically favourite value of $q_j$ from
such a complex potential?  One might have thought that $q_j$ should
minimise the real part of the potential.  Although such a working
hypothesis may give a practical prescription, this cannot be justified
from the first-principle approach.  This complex potential for the
Polyakov loop is one clear manifestation of the notorious sign problem
(see reference \cite{Muroya:2003qs} for an introductory review).

The sign problem hinders the lattice-QCD simulation at finite
density.  Contrary to what is believed, the sign problem is actually a
quite generic problem of the importance sampling not only in the
lattice-QCD simulation but in the mean-field approximation also
\cite{Dumitru:2005ng,Fukushima:2006uv}.  One should notice that the
mean-field approximation is based on the importance sampling; the
mean-field variables are chosen to be a ``configuration'' that
maximises the weight $\sim \rme^{-V_{\rm eff}[q]}$.  When the
potential is complex, therefore, the mean-field approximation breaks
down.

Occasionally, in some analytical studies, it is overemphasised that
the method be sign-problem free.  Such a statement must be misleading
as long as the method relies on the mean-field approximation for the
treatment of the gauge-field part such as the Polyakov loop dynamics.


\subsection{Non-perturbative methods in progress}

One of the most urgent challenges in theory is to outline the global
structure of the QCD phase diagram and fill in the quantitative
details on \fref{fig:minimal}.  Non-perturbative methods are
indispensable to access the information in the vicinity of phase
transition regions.  There are significant progresses recently in the
strong-coupling expansion, the large-$\Nc$ QCD and the effective
models to shed light on the phase diagram.  Unfortunately, not much
about the phase diagram can be said from the holographic QCD models,
though there are many interesting attempts on each state of QCD
matter, particularly by means of the AdS-Reissner-Nordstr\"{o}m black
hole \cite{Nakamura:2009tf,Nakamura:2006xk,Kim:2007em,Jo:2009xr},
where the charged black hole in AdS space is identified as the
finite-$T$ and finite-$\muq$ plasma.  There are some more
investigations on quark-matter and nuclear-matter properties using the
holographic approach \cite{Andreev:2010bv,Kim:2010dp} and also on the
phase transition in the Sakai-Sugimoto model
\cite{Sakai:2004cn,Aharony:2006da,Horigome:2006xu}.  In this article,
not only the holographic approaches but also the finite-density
studies by means of the orbifold equivalence
\cite{Cherman:2010jj,Hanada:2011ju} and the orientifold equivalence
\cite{Armoni:2003fb,Armoni:2003yv} are beyond our current
scope.


\subsubsection{Strong-coupling expansion and the matrix model:}

The Polyakov loop matrix model emerges as a result of the strong
coupling expansion and this model provides us with an ideal setup to
think of the sign problem.  In the leading order of the hopping
parameter expansion \cite{Green:1983sd}, in the presence of heavy
quarks, the quark-loop contribution or the Dirac determinant amounts
to
\begin{equation}
 \Md(\muq) \approx 1 + h \sum_{\bx} \biggl[ \rme^{\beta\muq}
  \tr L(\bx) + \rme^{-\beta\muq}\tr L^\dagger(\bx) \biggr]
\label{eq:strong-quark}
\end{equation}
with a coefficient $h$ which is small for large quark mass $\mq$ and
eventually $\Md(\muq)\to 1$ as $\mq\to\infty$ (quenched limit).
Because the Polyakov loop changes non-trivially under a centre
transformation, the above quark contribution breaks centre symmetry
explicitly.  Besides, this action becomes complex when a finite $\muq$
is turned on since $\tr L$ takes a complex value in general.

The expression~\eref{eq:strong-quark} is very useful to grasp the
nature of the sign problem.  One can also use
equation~\eref{eq:strong-quark} the other way around to deduce special
situations where the sign problem weakens.  We shall enumerate some of
widely acknowledged examples here:

\begin{itemize}
\item Two-colour QCD \cite{Nakamura:1984uz,Kogut:1999iv,Hands:2006ve}~ --- ~
If the colour gauge group is SU(2), the Polyakov loop is always real;
$\tr L = \tr L^\dagger = 2\cos(\pi q)$.
Equation~\eref{eq:strong-quark} is thus real.  Taking a real value is
not sufficient for feasibility of the importance sampling because the
Dirac determinant could be real but negative.  There must be an even
number of degenerate quark species in order to guarantee the
semi-positivity of the Dirac determinant.

\item Isospin Chemical Potential \cite{Son:2000xc,Kogut:2002zg}~ --- ~
We see that $\Md(-\muq)=\Md(\muq)^\ast$ from
equation~\eref{eq:strong-quark}, meaning that the whole Dirac
determinant is positive semi-definite if there are two degenerate
quarks that have a chemical potential opposite to each other.  For
example, $\mu_u=\muI$ for $u$-quarks and $\mu_d=-\muI$ for $d$-quarks
with $m_u=m_d$.  It is then easy to confirm that the Dirac determinant
satisfies,
\begin{equation}
 \Md(\mu_u)\Md(\mu_d) = |\Md(\muI)|^2 \ge 0 \;.
\end{equation}
The isospin chemical potential causes no sign problem, therefore, and
the Monte-Carlo simulations are feasible.  The absence of the sign
problem for the chiral chemical potential $\mu_5$ is also a variant of
this category \cite{Fukushima:2008xe}, for which some lattice
simulations are successful \cite{Yamamoto:2011gk,Yamamoto:2011qa}.

\item Imaginary Chemical Potential
  \cite{Alford:1998sd,deForcrand:2002ci,D'Elia:2009tm}~ --- ~
The sign problem originates from the imbalance between the quark and
anti-quark propagation in equation~\eref{eq:strong-quark}, which can
be made balanced by replacing $\muq$ by a pure-imaginary quantity
$\rmi \muIm$.  Then $\Md(\muIm)$ is obviously real.  In this case,
unlike two-colour QCD, there needs not be an even number of degenerate
quark species because $\rme^{\rmi\beta\muIm}$ is bounded.  (The
situation about the positivity of the Dirac determinant is rather
similar to the zero-density case.)   One may have thought that the
Dirac determinant is then a periodic function of $\muIm$ with a period
$2\pi T$.  This is not correct.  In fact the phase factor
$\rme^{\rmi\beta\muIm}$ can be partially cancelled by the centre
transformation and the rest takes a value from $1$ to
$\rme^{2\rmi\pi/\Nc}$.  This means that the genuine period is
$2\pi T/\Nc$ instead of $2\pi$ (i.e.\ Roberge-Weiss periodicity 
\cite{Roberge:1986mm}).
\end{itemize}

Even though the sign problem is not washed away, the mean-field
approximation works anyway (see \sref{sec:models} for details).  In
the strong-coupling expansion the gauge action is dropped and the
quark sector is dominant.  Then, it is not the deconfinement
transition but the chiral phase transition that defines the phase
diagram.  The phase boundary of chiral restoration obtained by means
of the staggered formalism of chiral fermions
\cite{Fukushima:2003vi,Nishida:2003fb,Miura:2009nu} is in qualitative
agreement with \fref{fig:minimal}, which has been also confirmed
by the numerical simulation \cite{deForcrand:2009dh}.

The effect of the Polyakov loop dynamics as formulated by the
action~\eref{eq:pol-matrix} has been taken into account too
\cite{Ilgenfritz:1984ff,Gocksch:1984yk,Fukushima:2002ew,Fukushima:2003fm}.
Such a treatment on the lattice can be easily translated into the
continuum language, which has led to the so-called PNJL-type models as
we discuss soon below.


\subsubsection{Large-$\Nc$ QCD:}
\label{sec:largeNc-density}

\begin{figure}
\begin{center}
\includegraphics[width=0.4\textwidth]{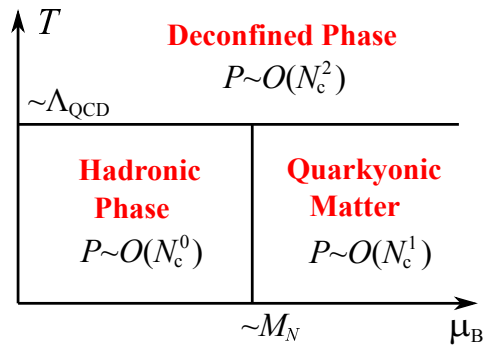}
\caption{The phase diagram of large-$\Nc$ QCD \cite{McLerran:2007qj}.
  The hadronic phase confines coloured excitations and only glueballs
  and mesons exist.  The corresponding system pressure is of
  $O(\Nc^0)$.  Above the deconfinement temperature of order of
  $\LQCD$, thermal excitations are dominated by gluons and the
  associated pressure is of $O(\Nc^2)$.  The quarkyonic phase is
  characterised by the pressure of $O(\Nc)$ at $\muB$ greater than the
  lightest baryon mass $M_N$ so that the baryon number density is
  non-vanishing.}
\label{fig:quarkyonic}
\end{center}
\end{figure}

Recently an interesting possibility about a new structure on the QCD
phase diagram has been suggested from analytic deliberations on the
large-$\Nc$ limit of QCD at finite $T$ and $\muB$
\cite{McLerran:2007qj}.  The phase diagram takes a simple structure as
sketched in \fref{fig:quarkyonic} with three regions separated
by straight first-order phase boundaries.

In the large-$\Nc$ limit quark loops are suppressed by $1/\Nc$ as
compared to gluon loops, so $\muB$ does not affect the deconfinement
transition line which is predominantly determined by $\Nc^2-1$ gluons.
Hence the deconfinement transition makes a straight line parallel to
the $\muB$ axis.  A free quark gas would give a pressure of
$\sim \Nc\muq^4\sim \Nc^{-3}\muB^4$, and thus the deconfinement
transition line would be deformed for $\muB$ as large as
$O(\Nc^{5/4})$ that surpasses the gluon degrees of freedom.  Before
this is reached, there appears another type of transition at
$\muB\simeq M_N$ where $M_N$ is the lightest baryon excitation energy
(baryon mass minus binding energy).  The baryon number density becomes
non-vanishing then and the system pressure jumps from $O(\Nc^0)$ in
the hadronic phase to $O(\Nc)$ in the new phase which is called
Quarkyonic Matter.  Before the deconfinement phase transition takes
place with $T\sim \LQCD$, glueballs and mesons cannot affect this
threshold for the baryon number and so the threshold located at
$\muB\simeq M_N$ makes a straight line parallel to the $T$ axis, which
results in the phase structure presented in \fref{fig:quarkyonic}.

The reason why the right-bottom region of \fref{fig:quarkyonic} is
identified as the Quarkyonic Matter is the following.  As explained
before, gluons and quarks are all confined below the deconfinement
transition line, and so the Quarkyonic Matter region resides in the
confined regime.  Therefore the physical degrees of freedom there
should be baryons rather than quarks, and one can prove that the dense
baryonic system at large $\Nc$ indeed gives the pressure of $\sim
O(\Nc)$ whose major contribution comes from not the Fermi energy but
the baryon-baryon interaction energy \cite{McLerran:2007qj}.  On the
other hand, a gas of quarks naturally yields a pressure of $O(\Nc)$
because of the presence of $\Nc$ quarks.  Such coincidence in the
$\Nc$ counting implies that this bottom-right state would confine
quarks and nevertheless feel quarks somehow.  In other words a dual
interpretation is possible;  matter of strongly-interacting baryons
and simultaneously that of weakly-interacting quarks, which motivated
the name, Quarkyonic Matter $=$ (Quark + Baryonic) Matter.

Such an interpretation may sound peculiar but there is a reasonable
way to reconcile two interpretations in terms of baryons and quarks.
That is, particles sitting deeply inside of the Fermi sea cannot be a
part of excitation spectra, and so they could be quarks even though
the system is in the confined phase.  Therefore, the Fermi sphere
consists of both baryons and quarks;  baryons in the momentum layer
$\sim \LQCD$ near the Fermi surface (but, because of the confining
interaction, there is no sharp Fermi surface in reality) and quarks
inside of the Fermi sphere which does not take part in the excitation
but gives the pressure of $\sim O(\Nc)$.

This is actually what should be expected in the large-$\Nc$ limit.  As
we have discussed in the beginning of this section, a large $\muq$ does
not guarantee the smallness of $\as(\muq)$, but the screening effects
due to quark polarisation make $\as(\mu)$ effectively small enough to
realise the perturbative regime at high density.  Because the quark
loops are suppressed in the $\Nc$ counting, the screening effects
would diminish and soft-gluon exchange would become important for
large $\Nc$.

The soft-gluon exchange would lead to confinement for excitations on
top of the Fermi surface, and furthermore, to an interesting
consequence for the chiral-symmetry breaking mechanism
\cite{Kojo:2009ha,Kojo:2011cn}.  Usually the homogeneous chiral
condensate $\langle \bar{q}q\rangle$ or the condensation of the sigma
meson at rest is attributed to the spontaneous breaking of chiral
symmetry.  Because quarks and quark-holes must sit near the Fermi
surface, if the net momentum of $\bar{q}q$-system is zero on the one
hand, the gluon exchanged between a quark and a quark-hole on the
Fermi surface must carry as large momentum as $2\muq$.  On the other
hand, this momentum can be absorbed in the net momentum $p\sim 2\muq$
of $\bar{q}q$-system and then the possible soft-gluon exchange favours
an inhomogeneous chiral condensate, namely, the chiral spiral state.

It is a subtle question whether the chiral spiral structure emerges
and Quarkyonic Matter exists in the real world at $\Nc=3$.  This
depends on the competition of the strength of confining force and the
screening due to quark polarisation.  For the sake of such
quantitative clarification the effective model study was expected to
hint the relevance or irrelevance of the large-$\Nc$ limit to the real
world.  It has turned out, however, that one should inevitably go
beyond the mean-field approximation and it is still a very difficult
question how to resum the Polyakov loop fluctuations.


\subsection{Extrapolation from effective models}
\label{sec:models}

The Polyakov loop behaviour in the pure gluonic sector is described
nicely by the simple parametrisation~\eref{eq:Pol-pot}.  The quark
loop on top of the Polyakov loop background in
equation~\eref{eq:Weiss-density-pre} is expressed in a gauge
invariant way, i.e.\
\begin{equation}
 \hspace{-3.5em}
 V_{\rm quark}^{(1)} = -2\Nf VT\int\frac{\rmd^3 p}{(2\pi)^3}
  \Bigl\{ \tr\ln\bigl[ 1+ L\, \rme^{-\beta(\epsilon-\muq)} \bigr]
  +\tr\ln\bigl[ 1+ L^\dagger\, \rme^{-\beta(\epsilon+\muq)} \bigr]
  \Bigr\} \;,
\label{eq:Vtr}
\end{equation}
where the quark mass is included in the energy dispersion relation
$\epsilon=\sqrt{p^2+\Mq^2}$ with the dynamically generated mass $\Mq$
which could be considerably larger than $\muq$.  This form of the
coupling is simple but has rich contents.  If chiral symmetry is badly
broken by large $\Mq$, the exponential terms are small and thus the
Polyakov loop coupling to the chiral sector diminishes, which is in
favour of confinement.  If the Polyakov-loop expectation value is
small, on the other hand, the thermal excitation of quarks is severely
screened and thus chiral restoration is delayed.

Interestingly enough, the colour trace in equation~\eref{eq:Vtr} is
explicitly taken to be a form of
\begin{eqnarray}
\hspace{-2em}
 V_{\rm quark}^{(1)} &= -2\Nf VT\int\frac{\rmd^3 p}{(2\pi)^3}
  \Bigl\{ \ln\Bigl[ 1 + \tr L \,\rme^{-\beta(\epsilon-\muq)}
  +\tr L^\dagger\, \rme^{-2\beta(\epsilon-\muq)}
  +\rme^{-3\beta(\epsilon-\muq)} \Bigr] \nonumber\\
 &\qquad\qquad\quad + \ln\Bigl[ 1
  +\tr L^\dagger \,\rme^{-\beta(\epsilon+\muq)}
  +\tr L\, \rme^{-2\beta(\epsilon+\muq)}
  +\rme^{-3\beta(\epsilon+\muq)} \Bigr] \Bigr\} \;.
\label{eq:PNJL-expand}
\end{eqnarray}
The further missing piece in the dynamics is the spontaneous breaking
of chiral symmetry now that the deconfinement and coupling parts are
formulated as explained above.  The Polyakov-loop coupled
Nambu--Jona-Lasinio model (PNJL model)
\cite{Fukushima:2003fw,Ratti:2005jh} utilises the NJL model as a
dynamical theory to describe the chiral condensate
$\langle\bar{q}q\rangle$.  Replacing the NJL model by another
successful chiral model, the Quark-Meson (QM) model (that is a variant
of the linear sigma model), one can define the PQM model
\cite{Schaefer:2007pw}.  In the simple mean-field approximation the
PNJL model is more convenient than the PQM model because the linear
sigma model suffers from artificial first-order phase transition
\cite{Baym:1977qb}.  This model artifact is cured by pion loop effects
implemented by the renormalisation group (RG) improvement.  For this
purpose of RG studies the PQM setup has an advantage over the PNJL
model \cite{Herbst:2010rf,Skokov:2010wb}.

So far the RG improvement on the meson fluctuations has been well
investigated, while it is not known how to include the Polyakov loop
fluctuations systematically.  Once the Polyakov loop potential is
given, there is no way to improve it, and a more fundamental starting
point is necessary.  One possibility is to take the Polyakov loop
matrix model that can in principle encompass soft fluctuations of the
Polyakov loop.  Since this matrix model is defined on the lattice,
however, it is technically difficult to accomplish the RG analysis.
Another possibility is to postulate a derivative term in the Polyakov
loop action, i.e.\ setup of the so-called Polyakov loop model
\cite{Dumitru:2000in} and to put it in the RG equation
\cite{Wetterich:1992yh}.

Here, let us make a brief remark on the sign problem in the PNJL-type
models.  It may seem to be sign-problem free in
equation~\eref{eq:PNJL-expand} but it is not so.  In the simple
mean-field analysis equation~\eref{eq:PNJL-expand} leads to
\begin{eqnarray}
\hspace{-4em}
 V_{\rm quark}^{(1)}[\Phi,\bar{\Phi}] &= -2\Nf VT\int\frac{\rmd^3 p}{(2\pi)^3}
  \Bigl\{ \ln\Bigl[ 1 + 3\Phi \,\rme^{-\beta(\epsilon-\muq)}
  +3\bar{\Phi} \, \rme^{-2\beta(\epsilon-\muq)}
  +\rme^{-3\beta(\epsilon-\muq)} \Bigr] \nonumber\\
 &\qquad\qquad\quad + \ln\Bigl[ 1
  +3\bar{\Phi} \,\rme^{-\beta(\epsilon+\muq)}
  +3\Phi \, \rme^{-2\beta(\epsilon+\muq)}
  +\rme^{-3\beta(\epsilon+\muq)} \Bigr] \Bigr\} \;.
\label{eq:pol_approx}
\end{eqnarray}
Here, at finite $\muq$, it is necessary to treat $\Phi$ and
$\bar{\Phi}$ independently.  Then, one can plot
$V_{\rm quark}^{(1)}[\Phi,\bar{\Phi}]$ as a function of $\Phi$ and
$\bar{\Phi}$ to recognise a funny shape.  In particular one may find
that the solution of the gap equation is not stable in the direction
of $\bar{\Phi}-\Phi$.  This is how the sign problem remains unsolved
in the mean-field model study and in fact identifying the saddle-point
as the ground state leads to an approximation similar to the
reweighting method to evade the sign problem \cite{Fukushima:2006uv}.

Let us make a brief comment on the approximation:
$\tr L\to 3\Phi$ in the logarithm in equation~\eref{eq:pol_approx}.
Such a replacement is acceptable only when the Polyakov loop
fluctuations are negligible as compared to its mean value $\Phi$.  To go
beyond the mean-field level the group integration with respect to the
Polyakov loop matrix $L$ should be incorporated as done in
\cite{Gocksch:1984yk,Fukushima:2002ew,Fukushima:2003fm,Megias:2004hj}.
Such a treatment of the group integration with respect to $L$ is
important to maintain gauge invariance;  in the literature
gauge-variant quantities such as the phases of the Polyakov loop are
chosen as the mean-field variables to simplify calculations with
colour-superconducting gaps, but they would lead to unphysical colour
density.  This problem is often neglected but must be resolved by the
group integration as elucidated in \cite{Abuki:2009dt}.

\begin{figure}
\begin{center}
\includegraphics[width=0.45\textwidth]{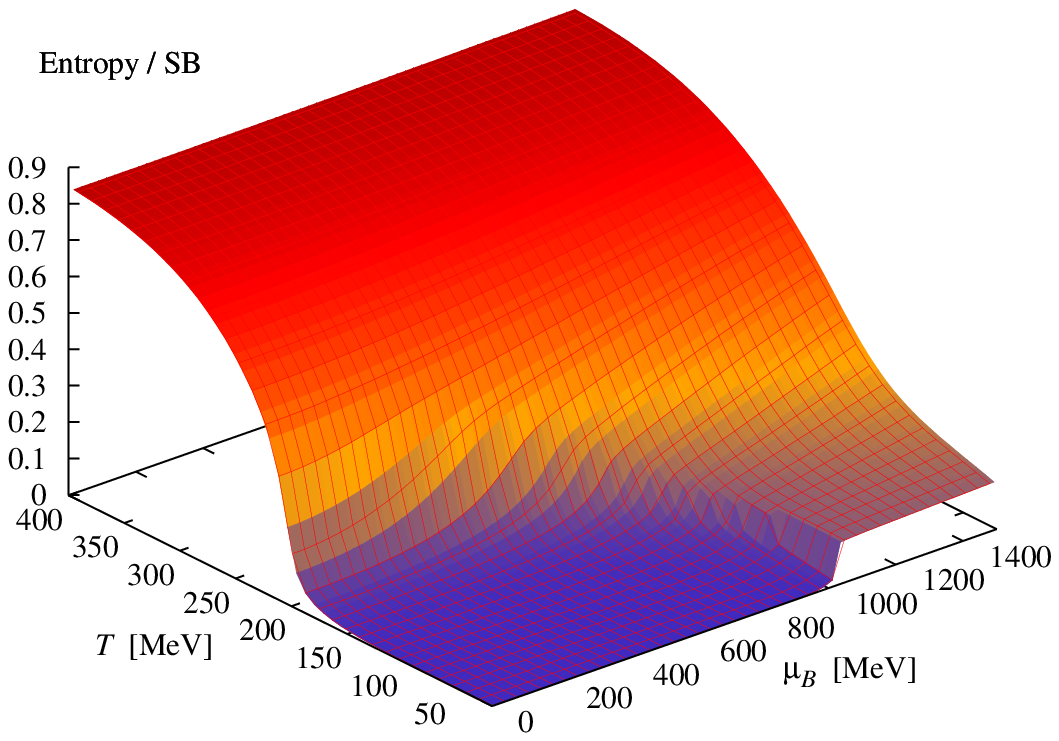} \hspace{1em}
\includegraphics[width=0.45\textwidth]{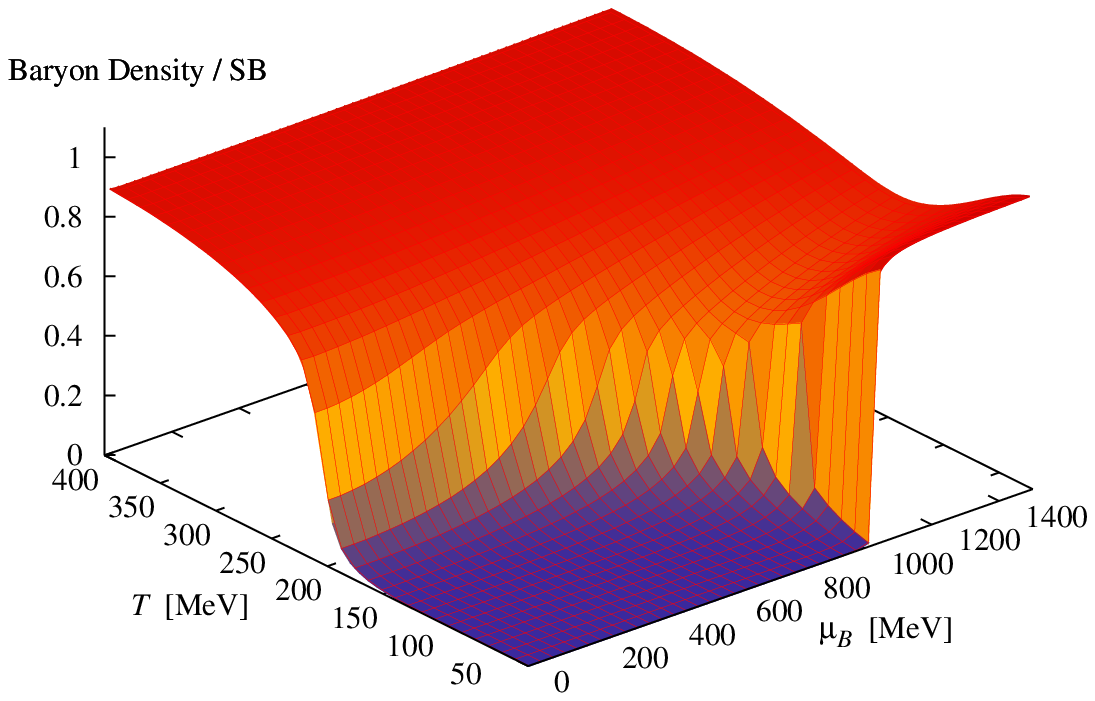}
\end{center}
\caption{Entropy density and the baryon number density normalised by
  the Stefan-Boltzmann value in the mean-field approximation of the
  PNJL model.  The entropy density, which is a $T$-derivative of the
  thermodynamic potential, should give information on gluon
  deconfinement, whereas the baryon number density, which is a
  $\muB$-derivative, should be sensitive to quark deconfinement.}
\label{fig:PNJL}
\end{figure}

Nevertheless, once an approximation is made with a prescription to
neglect the sign problem, the model results are useful to figure the
thermodynamic quantities out at finite $T$ and $\muB$.
Figure~\ref{fig:PNJL} shows some examples from the PNJL model in the
simple mean-field approximation \cite{Fukushima:2008wg}.  The left
is the entropy density divided by the Stefan-Boltzmann value and the
right is the baryon number density divided by the Stefan-Boltzmann
value.  Naturally the increase in the entropy density is to be
interpreted as deconfinement.  One may well conclude that the model
results could have implied the realisation of Quarkyonic Matter at low
$T$ and high $\muB$ where the entropy density stays small and the
baryon density gets large which is characteristic to Quarkyonic
Matter.

\begin{figure}
\begin{center}
\includegraphics[width=0.25\textwidth]{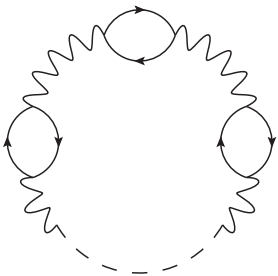}
\end{center}
\caption{Ring diagram of gluons with the quark pair
  creation/annihilation.  The Polyakov loop potential must have
  coupling to the quark chemical potential $\muq$ through this quark
  polarisation.  If an external $\bB$ is applied, gluons can feel
  $\bB$ through this diagram too.}
\label{fig:ring}
\end{figure}

However, the model study has missing diagrams, as depicted in
\fref{fig:ring}, which is not included even in the RG improvement and
is related to the Polyakov loop fluctuations.  This missing
contribution is critically important to clarify the QCD phase
structure including the possibility of Quarkyonic Matter.  Without
this diagram the Polyakov loop potential \eref{eq:Pol-pot} has no
explicit dependence on $\muq$, so that the deconfinement transition is
almost insensitive to $\muq$ even with the coupling effects through
equation~\eref{eq:Vtr}.  This approximate $\muq$-independence is
observed in the entropy behaviour too in the left of \fref{fig:PNJL}.
Then, the phase diagram from the PNJL or PQM models turns out similar
to the large-$\Nc$ conjecture in \fref{fig:quarkyonic}.  In fact, it
is the polarisation effect in \fref{fig:ring} that would make a
difference from the large-$\Nc$ limit.

It is not straightforward to implement the polarisation effects
properly in the effective model without losing simplicity.  The model
is to be appreciated as long as it is simple enough to deepen the
intuition of physics understanding.  There are only a few attempts to
incorporate effects originating from the diagram in \fref{fig:ring}
into the PNJL-type models.  By the hypothetical matching condition for
the deconfinement and chiral transitions the $\muq$-dependence was
introduced in reference \cite{Schaefer:2007pw}.  Then, the
deconfinement line comes along with the chiral restoration line on the
phase diagram, but this statement is a consequence by construction of
the model.  In reference \cite{Fukushima:2010is} the $\muq$-dependence
in the Polyakov loop potential was determined by the matching
condition to thermodynamics from the thermal Statistical Model, which
has confirmed quantitative agreement with the prescription in
reference \cite{Schaefer:2007pw}.

This kind of polarisation effect is of increasing importance in the
researches on finite-density QCD matter.  Besides, as we see in the
next section, QCD in strong magnetic fields is currently a hot subject
and it requires detailed information on the polarisation effect.  The
reason for this is exactly the same as the finite-density case.
Gluons do not feel the magnetic fields directly, but do see them
through the quark polarisation as in \fref{fig:ring}.


\section{Strong Magnetic Field and Dimensional Reduction}
\label{sec:magnetic}

\begin{figure}
\begin{center}
\raisebox{2ex}{\includegraphics[width=0.35\textwidth]{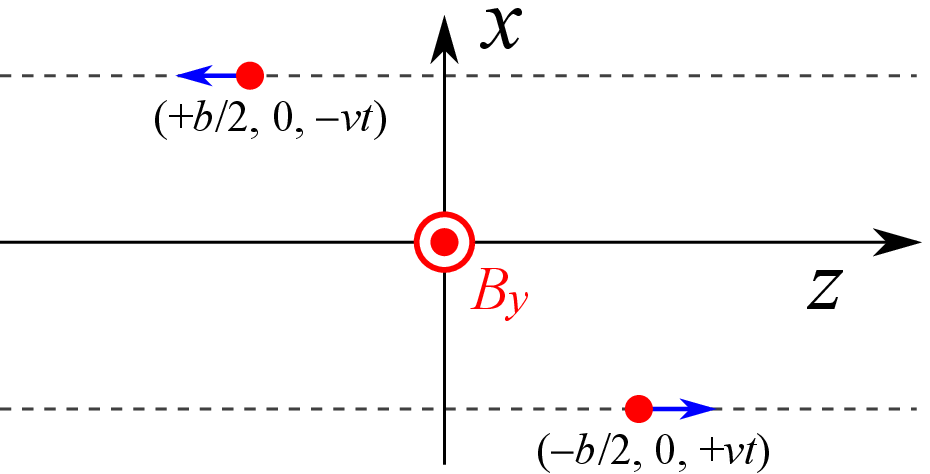}}
\hspace{5ex}
\raisebox{-2ex}{\includegraphics[width=0.35\textwidth]{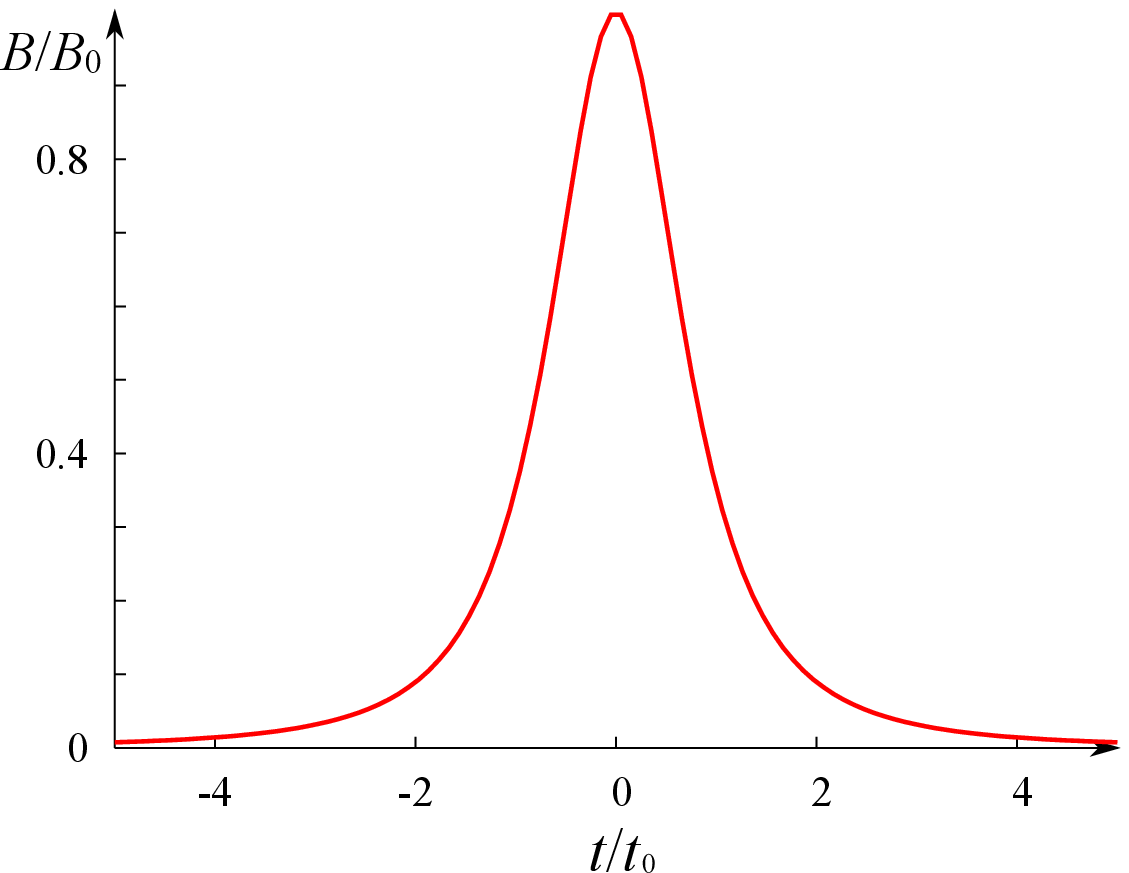}}
\end{center}
\caption{Left: Collision geometry seen from the above.\ \ Right:
  Profile of the produced magnetic field as a function of time.}
\label{fig:mag}
\end{figure}

In the heavy-ion collision with a finite impact parameter
(i.e.\ peripheral collision) a magnetic field is created by the
positively charged ions moving at almost the speed of light.  Let us
evaluate how large magnetic field is expected in the collision at the
RHIC energy in a classical manner.  For simplicity we assume that the
(positively charged) heavy ions are point charges
\cite{Kharzeev:2009pj}.  The collision geometry is schematically
modelled as in the left of \fref{fig:mag}.  Then, from the
Li\'{e}nard-Wiechert potential, the magnetic fields at the origin
reads
\begin{eqnarray}
 & e\bB(\bx,t) = \frac{Ze^2}{4\pi}\cdot\frac{b\beta(1-\beta^2) \bi{e}_y}
  {[(\beta t)^2 + (1-\beta^2)(b/2)^2 ]^{3/2}}
  = eB_0 \frac{\bi{e}_y}{[1+(t/t_0)^2]^{3/2}} \;,\\
 & eB_0 = \frac{8Z \alpha_e}{b^2} \sinh(Y)
  = (47.6\MeV)^2 \Bigl(\frac{\mbox{1fm}}{b}\Bigr)^2
  Z \sinh(Y) \;,\nonumber\\
 & t_0 = \frac{b}{2\sinh(Y)} \;.\nonumber
\end{eqnarray}
In the definition of $B_0$ and $t_0$ we use the beam rapidity $Y$
instead of the velocity $\beta$, which is related by
$\beta=\tanh(Y)$.  Here, $B_0$ is the maximum strength of the magnetic
field and $t_0$ gives a typical time scale of decaying field.  In
the case of Au-Au collision at the RHIC energy, these parameters are
\begin{equation}
 Z = 79\;, \qquad \sinh(Y)\simeq \cosh(Y) =
  \frac{\sqrt{s_{_{NN}}}}{m_N} \simeq 106.6\;.
\end{equation}
The point-charge approximation is valid when the collision is far
peripheral.  So, let us take $b=10\fm$ \cite{Kharzeev:2009pj}.  Then,
this simple estimate leads to
\begin{equation}
 eB_0 \simeq 1.9\times 10^5\MeV^2 = 3.2\times 10^{19}\mbox{gauss} \;,
  \qquad t_0 \simeq 0.05\fm/c \;.
\end{equation}
This magnetic field strength is $10^4$ times larger than the surface
magnetic field of the magnetar, and $10^7$ times larger than that of
the ordinary neutron star.  Although such a strong field is transient
and decays with the time scale $t_0$, we note that the decay is not as
steep as exponential damp but power-law suppression.
At $t/t_0\sim 2\sim 0.1\fm/c$, for example, the magnetic field
diminishes to a tenth of $B_0$.  We note that this time scale is of
order of $Q_s^{-1}$ where $Q_s$ is the saturation scale at RHIC
\cite{McLerran:1993ni,McLerran:1993ka,McLerran:1994vd}.  Although
there are a number of theoretical calculations of equilibrated QCD
matter under strong $\bB$ fields, any serious simulation of the Glasma
\cite{Lappi:2006fp} and the particle production in strong $\bB$ have
not been fully analysed.  The real-time dynamics of the strong $\bB$
effects needs more investigations.


\subsection{Topological properties probed by the magnetic field}

It is well-known that special gauge configurations with non-zero
winding number play an important role in understanding of the vacuum
structure in the strong interactions.  The spontaneous breaking of
chiral symmetry is attributed to the QCD instanton which is the origin
of dynamical mass generation \cite{Diakonov:1995ea}.  The confinement
nature is also explained in terms of magnetic monopole condensation in
a special class of the gauge choice (see \cite{'tHooft:1999au} for a
lecture note).

There is no doubt about the existence of topological configurations in
QCD physics, but it is quite challenging how to ``see'' such
topological contents in real experiments.  The Chiral Magnetic Effect
(CME) is one of the promising candidates
\cite{Fukushima:2008xe,Kharzeev:2007jp,Kharzeev:2009fn}.  Let us
imagine the following situation specifically;  the QCD vacuum
accommodates one instanton that has a topological charge $Q_W$ and a
magnetic field $\bB$ that is as strong as the QCD energy scale $\LQCD$
is applied on the instanton.

Then, the axial anomaly relation (for the single-flavour case),
\begin{equation}
 \partial_\mu j_5^\mu = -\frac{g^2}{8\pi^2}\int d^3x \,\tr
  F_{\mu\nu}\widetilde{F}^{\mu\nu} \;,
\label{eq:anomaly}
\end{equation}
implies that
\begin{equation}
 \Delta N_5 = N_5(t=\infty) - N_5(t=-\infty) = -2Q_W \;.
\label{eq:index}
\end{equation}
The topological charge $Q_W$ is given by the temporal integration of
the right-hand side of equation~\eref{eq:anomaly}.  This means that,
if the system starts with the chirally neutral situation
($N_5(t=-\infty)=0$), a finite amount of chirality in the final state
is generated by the topological charge during the time evolution.  In
the chiral limit the momentum and the spin are parallel to each other
if the chirality is right-handed, while they are anti-parallel if the
chirality is left-handed.  The spin is aligned by strong $\bB$, which
makes the momentum also aligned along the $\bB$ direction, leading to
a non-vanishing value of the total momentum if $\Delta N_5\neq0$.  In
other words, since Dirac fermions are charged, an electric or baryonic
current is produced for $\bB\neq0$ and $\Delta N_5\neq0$.

\begin{figure}
\begin{center}
 \includegraphics[width=0.2\textwidth]{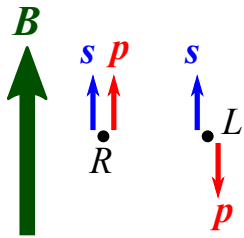}
\end{center}
\caption{Chiral Magnetic Effect:  In strong $\bB$ field right-handed
  particles move in parallel with $\bB$ and left-handed particles in
  anti-parallel to $\bB$.  A non-zero net flow results from
  $N_5\neq0$.}
\label{fig:cme}
\end{figure}

Such an effect can be expressed simply as \cite{Kharzeev:2007jp}
\begin{equation}
 \bJ_V = -2Q_W \frac{\bB}{|B|} \;,
\label{eq:cmeq}
\end{equation}
for large enough $\bB$, where $\bJ_V$ represents the vector current
$\langle\bar{\psi}\gamma^\mu\psi\rangle$ integrated over the spatial
volume.

For arbitrary strength of $\bB$ it is more appropriate to work in the
grand canonical ensemble using the chiral chemical potential $\mu_5$
instead of $N_5$.  One can then prove non-perturbatively that
\cite{Fukushima:2008xe,Metlitski:2005pr,Vilenkin:1980fu}
\begin{equation}
 \bj_V = \frac{e \mu_5}{2\pi^2}\bB
\label{eq:cme}
\end{equation}
holds for any $\bB$ and $\mu_5$.  It is easy to confirm that
equation~\eref{eq:cme} is reduced to equation~\eref{eq:cmeq} in the
strong $\bB$ limit using the anomaly relation~\eref{eq:index}.

In fact this is one example of more generic relation,
\begin{equation}
 j^\mu = C \varepsilon^{\mu\nu\rho\sigma}\partial_\nu\phi\, F_{\rho\sigma} \;.
\label{eq:cme-gen}
\end{equation}
Here $\phi$ is some field, and in the context of the CME, the strong
$\theta$-angle is identified as $\phi$ in the above.  Then, $\nu=0$,
$\rho=x$ and $\sigma=y$ uniquely fix $\nu=z$ from
$\varepsilon^{\mu\nu\rho\sigma}$.  By regarding $\partial_0\theta$ as
$\mu_5$ \cite{Fukushima:2008xe}, one can readily retrieve
equation~\eref{eq:cme} apart from the overall constant $C$.  One can
also understand related effects from equation~\eref{eq:cme-gen}.  If
$\theta$ (or $\phi$ in the above) is spatially inhomogeneous,
$\nu=z$ (instead of $\nu=0$) leads to $\mu=0$ (instead of $\mu=3$),
meaning that an electric charge or an electric dipole moment is
induced \cite{Kharzeev:2007tn}.  In addition, one can even think of
not the $\theta$-angle but the pion field $\pi$ as $\phi$ in
equation~\eref{eq:cme-gen}, and then the presence of time-dependent
background, i.e.\ $\partial_0 \pi^0$, results in an effect similar to
the CME-induced current.  It is actually argued in the Skyrmion model
that an additional electric charge is induced in the baryon content
(that is constructed as a profile of $\pi^a(\bx)$) under a strong
$\bB$ field \cite{Eto:2011id}.

What is detectable in experiments should not be the current $\bj_V$
itself because the QCD vacuum has fluctuations of instantons and
anti-instantons.  In other words, the parity ($\calP$) and the
charge-parity ($\calCP$) symmetries are broken only (spatially and
temporally) locally at the topological excitation, but those
symmetries are restored on average over fluctuations.  Thus, the
CME-induced current $\bj_V$ is also a local object and the (ensemble
or spatial) average makes it vanishing.  In this way the CME is one of
the candidates to give an account for the ``Local Parity Violation''
(LPV), if any, that might be observed in the relativistic heavy-ion
collisions.

The most relevant quantity to experimental data \cite{:2009uh} is
fluctuation of the CME-induced current, that is, the electric-current
susceptibility $\chi_j$ \cite{Fukushima:2009ft}.  The one-loop result
in the small-frequency limit with the zero momentum limit first taken
is
\begin{equation}
 \chi_j = \frac{e^2 |eB|}{2\pi^2} \;,
\label{eq:chij}
\end{equation}
for QED loops ($e^2$ should have been $g^2$ for quark loops), which
does not depend on $\mu_5$ and comes from the Landau zero-mode alone,
interestingly.  There is an intuitive argument to take a short-cut for
the derivation of the above expression \cite{Fukushima:2009ft}.  For
this purpose let us consider the electric-current generation rate
under strong $\bE$ as well as $\bB$, that is,
\begin{equation}
 \frac{\rmd(eJ_V)}{\rmd t} = V \frac{e^2|eB|E}{2\pi^2} \;,
\label{eq:djdt}
\end{equation}
which originates from the correspondence between the chirality
generation and the particle production when fields are strong
enough \cite{Fukushima:2010vw}.  The same quantity can be expressed in
the framework of the linear response theory as
\begin{eqnarray}
 \frac{\rmd(eJ_V)}{\rmd t} &= -\int\rmd^3x\,\rmd^4x'\, \Bigl\langle
  \frac{\rmd(ej_V)(x)}{\rmd t} j_V(x') \Bigr\rangle
  \,eA(x') \nonumber\\
 &= \int\rmd^3x\,\rmd^4x'\, e^2\langle j_V(x) j_V(x') \rangle\, E \;,
\label{eq:djdt2}
\end{eqnarray}
where $A(x)$ denotes a vector potential component parallel to
$\bB$ and $\bJ_V$.  From the first line to the second line above, we
used $E=\partial_0 A$.  By equating these \eref{eq:djdt} and
\eref{eq:djdt2}, we can immediately find $\chi_j$ given by
equation~\eref{eq:chij}.


\subsection{Implication to and from the QCD phase transitions}

Once the CME is confirmed in the heavy-ion collision experiment, it
would signal chiral symmetry restoration.  This is because, as we have
seen in the previous subsection, the CME requires massless Dirac
fermions and thus vanishingly small chiral condensate.  It is not
obvious what the CME can imply for deconfinement.  Generally speaking,
the sphaleron transition rate is proportional to $T^4$ by dimensional
reason and real-time topological excitations become abundant at higher
temperature \cite{Arnold:1987zg,McLerran:1990de,Bodeker:1999gx}.  In
this argument, however, any feature inherent to deconfined gluons and
quarks is not quite needed for the manifestation of the CME.\ \  In a
solvable model in (1+1) dimensions, as we will discuss later in fact,
the CME exists even though no deconfinement takes place.

In the computation of the electric-current susceptibility there is a
significant influence from the chiral phase transition.  Actually, at
finite $\mu_5$, the divergent chiral susceptibility has a mixing with
the current susceptibility $\chi_j$ causing enhancement in $\chi_j$ at
the chiral phase transition \cite{Fukushima:2010fe}.  It is not yet
understood how the Polyakov loop dynamics and the deconfinement
transition should affect $\chi_j$ and any other observable sensitive
to the CME.

In principle the lattice-QCD simulation in a strong $\bB$ field can
clarify the effect of chiral restoration and deconfinement
\cite{Yamamoto:2011gk,Yamamoto:2011qa,Abramczyk:2009gb,Buividovich:2009wi}.
There are also some effective model studies on the phase diagram
modified by $\bB$ and also the effects of the strong $\theta$-angle
\cite{Boer:2008ct,Mizher:2008hf,Boomsma:2009eh,Mizher:2010zb}.
It is known by now
that the PNJL and PQM model calculations with large $\bB$
\cite{Fraga:2008qn} are not
consistent with the lattice-QCD results in which chiral restoration
and deconfinement are locked together for any $\bB$
\cite{D'Elia:2010nq}.  This inconsistency could perhaps arise from the
missing diagram as shown in \fref{fig:ring} and the coupling to $\bB$
in the Polyakov loop potential.  Besides, the chiral model part may
have a non-trivial dependence on the Polyakov loop through the
fermionic interaction terms \cite{Kondo:2010ts,Gatto:2010pt}.  We also
note that dynamical locking mechanism as discussed in
\cite{Braun:2011fw} may play a role.  At least, it should be quite
robust that the chiral condensate is enhanced by $\bB$ due to the
magnetic catalysis \cite{Gusynin:1994re,Gusynin:1994xp} and thus the
chiral transition temperature is pushed up accordingly.

A careful comparison between results from the effective model approach
and from the lattice-QCD simulation at large $\bB$ should be very
useful for the finite-density study on QCD matter.  In the case at
$\muq>T$ it is difficult to impose any reliable constraint from the
lattice-QCD simulation, while the finite-$\bB$ simulation has no
principle problem and one can check quantitatively if the effective
model description is valid or inadequate.  If it is insufficient to
reproduce the lattice data at large $\bB$, it is most unlikely that
the same model can encompass the finite-density property of QCD matter
either.

One might have thought that the perturbative calculation of the Weiss
potential~\eref{eq:Weiss} can be extended to the finite-$\bB$ case.
It is just straightforward to generalise the quark one-loop
contribution~\eref{eq:Weiss-quark} to the finite-$\bB$ calculation.
What is more difficult is the polarisation effect as in
\fref{fig:ring}.  This has been evaluated in the Lowest Landau-Level
(LLL) approximation that will be explained in the next subsection for
a quick derivation of the CME-induced current and susceptibility.  In
short summary, the polarisation effect on the Weiss potential has the
following effect:  The Weiss potential~\eref{eq:Weiss} comes from the
integration with respect to two transverse gluons, which are physical
degrees of freedom and unphysical longitudinal and ghost modes cancel
out.  In the LLL approximation one can show that only one of two
transverse gluons acquires a screening mass proportional to
$\sqrt{eB}$.  The height of the potential, therefore, decreases up to
a half of the original Weiss potential with increasing $\bB$.
Qualitatively, a larger $\bB$ tends to reduce the barrier at the
confined state with $\Phi=0$, strengthen confinement and delay the
deconfinement phase transition.


\subsection{Dimensional reduction}

Under a strong magnetic field, in general, the transverse motion of
charged particles is equivalent to the one in the harmonic
oscillator.  The energy level is then discrete due to the Landau
quantisation.  Spin-$1/2$ fermions have the Landau zero-mode which
would dominate in the dynamics at energies below the scale
$\sim \sqrt{eB}$.   Such a strong $\bB$ enables us to use the LLL
approximation and to drop the transverse motion completely.  In this limit
we can reduce the (3+1)-dimensional theory into a form of the
(1+1)-dimensional one multiplied by the Landau level density.

In Minkowskian space-time we use the metric $g^{00}=-g^{11}=1$,
$g^{01}=g^{10}=0$ and the $2\times 2$ $\gamma$-matrices which satisfy
$\{\gamma^\mu,\gamma^\nu\}=2g^{\mu\nu}$.  Chirality is characterised
by $\gamma^5 = \gamma^0 \gamma^1 = \diag(1,-1)$ in the chiral (Weyl)
representation.  Therefore the upper (lower) element of two-component
spinor $\psi=(\psi_R,\psi_L)^t$ represents the right-handed
(left-handed) particle.  In (1+1) dimensions the
particle--anti-particle difference is correlated with the chirality.
That is, in momentum space, the right-handed component corresponds to
a right-moving ($p>0$) particle and a left-moving ($p<0$)
anti-particle.  One can understand the left-handed component in the
same way, i.e.\ a left-moving ($p<0$) particle and a right-moving
($p>0$) anti-particle.

In (1+1) dimensions the following relation among the $\gamma$-matrices
plays an interesting role for the topological currents;
\begin{equation}
 \gamma^\mu\gamma^5 = -\varepsilon^{\mu\nu}\gamma_\nu \;,
\label{eq:gamma_rel}
\end{equation}
where
$\varepsilon^{01}=-\varepsilon^{10}=-\varepsilon_{01}=\varepsilon_{10}=1$,
which relates the vector and the axial-vector currents.  As usual, we
can write the vector and the axial-vector currents as
\begin{equation}
 j_V^\mu = \bar{\psi}\gamma^\mu \psi \;,
 \qquad
 j_5^\mu = \bar{\psi}\gamma^\mu \gamma^5 \psi \;.
\end{equation}
Using the relation~\eref{eq:gamma_rel}, we have a relation,
$j_5^\mu=-\varepsilon^{\mu\nu} j_\nu$, that is explicitly written as
\cite{Basar:2010zd}
\begin{equation}
 j_V^1 = j_5^0 , \qquad j_5^1 = j^0 \;.
\end{equation}


\subsubsection{Topological currents in (1+1) dimensions:}

The relation between the vector and axial-vector currents is very
useful because, as we will see here, it captures the essential feature
of the CME-induced currents in (3+1) dimensions.

Let us consider the anomaly relation in (1+1) dimensions.  It is
well-known that the axial anomaly leads to
\begin{equation}
 \partial_\mu j_5^\mu = \frac{e}{2\pi}\epsilon^{\mu\nu} F_{\mu\nu}
 = \frac{e}{\pi} E = -2q_W \;,
\label{eq:anomaly2dim}
\end{equation}
where the electric field is $E=F^{10}$ in the standard convention.
Note that there is no magnetic field but only the electric field $E$
in (1+1) dimensions.  We here defined the (1+1)-dimensional
topological charge density as $q_W = -(e/2\pi)E$ in accord to the
convention.  By integrating equation~\eref{eq:anomaly2dim} over
space-time and assuming that the current falls sufficiently fast at
spatial infinity, one can recover equation~\eref{eq:index} easily.  We
can also prove that the topological charge,
$Q_W = \int \rmd^2x\, q_W(x)$, takes an integer number so that the
boundary condition in the $x$-direction can be maintained.

We also note that one can express equation~\eref{eq:anomaly2dim} in
the following form;
\begin{equation}
 \partial_\mu j_5^\mu(x) = -\frac{e}{\pi}
  (\partial^0 A^1 - \partial^1 A^0) = -2\partial_\mu K^\mu(x)
\end{equation}
with the (1+1)-dimensional Chern-Simons current density defined by
$K^\mu=-(e/2\pi) \varepsilon^{\mu\nu}A_\nu$.  From this
identification, the Chern-Simons number in this system is inferred as
\begin{equation}
 \nu(t) = \int \rmd x\, K^0(t,x)
  = \frac{e}{2\pi}\int \rmd x\, A^1(t,x) \;.
\end{equation}

Combining these expressions with the relation $j_V^1=j_5^0$ (where
$N_5$ is the volume integral of $j_5^0$), one can immediately write
the vector current integrated over space as
\begin{equation}
 J_V^1(t) = N_5(t) = -2\int_{-\infty}^t \rmd t'\,\rmd x\, q_W(t',x) \;,
\label{eq:cme_t}
\end{equation}
assuming that $N_5$ was zero at the initial time
($N_5(t=-\infty)=0$).  This simple relation leads to the current at
late time as given by
\begin{equation}
 J_V^1 = -2Q_W \;.
\end{equation}
This is nothing but the result expected when the spin is fully
polarised in the (3+1)-dimensional CME at strong $\bB$ (see
equation~\eref{eq:cmeq}).  Note that in (1+1) dimensions the spin is
always fully polarised because there is only one spatial direction and
thus the moving direction (either $p>0$ or $p<0$) and the chirality of
particles have one-to-one correspondence.  Here
equation~\eref{eq:cme_t} physically means Ohm's law because the
(1+1)-dimensional topological charge density is proportional to the
electric field as seen in equation~\eref{eq:anomaly2dim}.

If the spatial component of the Chern-Simons current falls
sufficiently fast, the topological charge is written as
$Q_W = \nu(t=\infty)-\nu(t=-\infty)$.  Therefore, (the spatial average
of) $A^1$ is the Chern-Simons number and the boundary condition of
$A^1$ in the $t$-direction gives the topological winding number.
Supposed that $\nu(t=-\infty)=N_5(t=-\infty)=0$, the topologically
induced current is written as
\begin{equation}
 J_V^1(t) = -\frac{e}{\pi}\int \rmd x\, A^1(t,x) \;.
\label{eq:JA1}
\end{equation}
If we identify $-eA^0$ as the chemical potential $\muq$ (regarding the
sign, remember the covariant derivative $p^0-eA^0$ and the dispersion
relation $p^0=E_p-\muq$ for particles).  Equation~\eref{eq:gamma_rel}
implies that $eA^1\gamma^1=eA^1\gamma^0\gamma^5$ and thus $-eA^1$ can
be regarded as the axial (or chiral) chemical potential $\mu_5$.
Therefore, one can reach a conclusion that
\begin{equation}
 J_V^1 = \frac{1}{\pi}\int \rmd x\, \mu_5 \;,
\label{eq:Jmu5}
\end{equation}
which correctly recovers the (3+1)-dimensional CME-induced
current~\eref{eq:cme} once this is multiplied by the Landau level
density, $eB/(2\pi)$.  That is,
\begin{eqnarray}
 j_V = \frac{\mu_5}{\pi}
  &\qquad \mbox{[in (1+1) dimensions]} \nonumber\\
 \longrightarrow \;
  j_V = \frac{|eB|}{2\pi} \cdot \frac{\mu_5}{\pi} \;,
  &\qquad \mbox{[in (3+1) dimensions]}
\end{eqnarray}
which coincides with equation~\eref{eq:cme}.

Here, it is clear that the longitudinal gauge field $A^1$, which is
the Chern-Simons number in (1+1) dimensions, plays the role of the
chiral chemical potential $\mu_5$ in (3+1) dimensions.  We note,
however, that there is an important difference;  usually $\mu_5$ is
introduced by hand as a constant, but in (1+1) dimensions $A^1$ must
have $t$-dependence to allow for nonzero $Q_W$.  We can think of a
concrete ``instanton'' configuration in (1+1) dimensions simply as
\begin{equation}
 A^1(t,x) = \frac{2\pi Q_W}{eL}\frac{t}{T} = -Et \;,
\end{equation}
where we limit ourselves to the spatially homogeneous case and denote
the spatial and temporal extents as $L$ and $T$, respectively, and
then we have
\begin{equation}
 j_V^1(t) = \frac{J_V^1(t)}{L} = \frac{e E}{\pi} t \;.
\label{eq:Jgen}
\end{equation}
From this, again, if multiplied by the Landau-level degeneracy we can
correctly recover the current generation rate given by
equation~\eref{eq:djdt}, i.e.
\begin{eqnarray}
 \frac{\rmd(ej_V)}{\rmd t} = \frac{e^2E}{\pi}
  &\qquad \mbox{[in (1+1) dimensions]} \nonumber\\
 \longrightarrow\;
  \frac{\rmd(ej_V)}{\rmd t} = \frac{|eB|}{2\pi}
  \cdot \frac{e^2E}{\pi} \;,
  &\qquad \mbox{[in (3+1) dimensions]}
\end{eqnarray}
which coincides with equation~\eref{eq:djdt}.

In the same way we can get a finite axial-vector current at finite
quark chemical potential $\muq$.  To see the anomalous nature in this
case the important fact is that the relation between the density and
the chemical potential is given by the quantum anomaly in (1+1)
dimensions, that is,
\begin{equation}
 n_{\rm q} = -\frac{eA^0}{\pi} \;,
\end{equation}
which results from the (1+1)-dimensional anomaly.  One can derive this
expression directly from $n=\langle\psi^\dagger(x)\psi(x)\rangle$ by
inserting the gauge field as a regulator as
$\lim_{y^0\to x^0}\psi^\dagger(y) \exp[-\rmi e\int \rmd t A^0]\psi(x)$.
From this one can immediately find,
\begin{equation}
 J_5^1 = \int \rmd x\, n_{\rm q} = \frac{1}{\pi}\int \rmd x\,\muq ,
\end{equation}
which represents the axial counterpart of the CME
\cite{Kharzeev:2007tn}.  This is again the anomaly relation exactly
same as that in (3+1) dimensions once multiplied by the Landau level
density $eB/2\pi$.


\subsubsection{Chiral Magnetic Effect in the Schwinger model:}

So far the arguments and the resulting expressions are quite general.
From now on we shall go into the dynamical properties calculating
microscopic quantities in a solvable (1+1)-dimensional model,
i.e.\ the massless Schwinger model.  The easiest way to accomplish a
calculation in the Schwinger model is to use the mapping onto a free
bosonic theory.  In our case, however, the bosonisation rule is a bit
more complicated than usual because we deal with not only fermionic
fields (such as the chiral condensate) but also gauge fields (such as
the electric field).  So, the Lagrangian density of the corresponding
theory should be
\begin{equation}
 \mathcal{L} = \frac{1}{2}(\partial^\mu \theta)(\partial_\mu \theta)
   -m_\gamma (\partial^\mu \theta)(\partial_\mu \phi)
   -\frac{1}{2}(\partial^\mu \phi)\partial^2
   (\partial_\mu \phi)
\label{eq:scalar}
\end{equation}
with the boson mass \cite{Schwinger:1962tp},
\begin{equation}
 m_\gamma^2 = \frac{e^2}{\pi} .
\end{equation}
After integrating the $\phi$-field out, we get a theory only in
terms of the $\theta$-field that is free (no interaction term) and has
a mass $m_\gamma$.  Such a scalar theory is usually used with the
bosonisation rule \cite{Smilga:1992hx},
\begin{eqnarray}
 & j_V^\mu = \bar{\psi}\gamma^\mu \psi
  = \frac{1}{\sqrt{\pi}} \varepsilon^{\mu\nu} \partial_\nu \theta \;,
\label{eq:boson_j}\\
 & j_5^\mu = \bar{\psi}\gamma^\mu\gamma^5 \psi
  = -\frac{1}{\sqrt{\pi}} \partial^\mu \theta \;,
\label{eq:boson_j5}\\
 & \bar{\psi}\psi = -c\, m_\gamma : \cos(2\sqrt{\pi}\theta) :
\end{eqnarray}
with the normal ordering $:\;:$.  Now we remark that $\phi$ in the
Lagrangian density~\eref{eq:scalar} comes from the gauge field,
$A^\mu = -\varepsilon^{\mu\nu} \partial_\nu\phi$ (where $\phi$
includes an instanton-like configuration $\sim \frac{1}{2}Et^2$ which
does not satisfy the periodic boundary condition in the
$t$-direction).  Then the electric field takes a form
$E=\partial^2\phi$.  Once we integrate the $\theta$-field out from the
theory, after the Gaussian integration in the functional formalism,
equation~\eref{eq:boson_j5} is replaced by
\begin{equation}
 j_5^\mu = -\frac{1}{\sqrt{\pi}}\partial^\mu \theta
  \;\;\to\;\; -\frac{m_\gamma}{\sqrt{\pi}} \partial^\mu \phi
  = -\frac{e}{\pi}\partial^\mu \phi \;.
\end{equation}
The anomaly relation is then derived as
\begin{equation}
 \partial_\mu j_5^\mu = -\frac{e}{\pi} \partial^2 \phi
  = -\frac{e}{\pi} E = -2q_W \;,
\end{equation}
which is fully consistent with the anomaly
relation~\eref{eq:anomaly2dim}.

In the same manner we can express the vector current in terms of
$\phi$ to find,
\begin{equation}
 j_V^\mu = \frac{e}{\pi}\varepsilon^{\mu\nu} \partial_\nu\phi
  = 2\varepsilon^{\mu\nu}\frac{\partial_\nu}{\partial^2} q_W \;.
\end{equation}
The inverse Laplacian should be understood in frequency and momentum
space (see equation~\eref{eq:jq}).  It is easy to make sure that this
result is fully consistent with the previous relation again.  That is,
after the spatial integration on $\phi$ and $q_W$ in the above, the
spatial derivative $\partial_1$ drops off and the right-hand side
simplifies as $-2/\partial_0$ for $\mu=1$ component, that is just the
$t$-integration.  Therefore the right-hand side finally becomes
$-2Q_W$ together with the spatial integration, and hence we obtain
$J_V^1=-2Q_W$.

The above equation gives a microscopic structure of the current in
more general cases with spatial modulation.  In frequency and momentum
space we can re-express this as follows;
\begin{equation}
 j_V^1(\omega,k)
  = \frac{-2\rmi \omega}{\omega^2 -k^2}\, q_W(\omega,k) \;.
\label{eq:jq}
\end{equation}
This is an interesting relation.  If $\omega\to 0$ is taken first, we
see that $j_V^1(0,k)$ is vanishing.  To get the CME-induced current
and the non-zero chiral magnetic conductivity, it is necessary to take
the zero-momentum limit in the order of $k\to0$ first and then
$\omega\to0$ later.  This observation is in fact consistent with the
result of the one-loop calculation of the conductivity
\cite{Kharzeev:2009pj}.

We point out that the structure of equation~\eref{eq:jq} naturally
appears from the transverse projection.  That is, after the one-loop
integration with the gauge potential source in momentum space, the
well-known result reads;
\begin{equation}
 j_V^\mu(\omega,k) = -\Biggl( g^{\mu\nu}-\frac{q^\mu q^\nu}{q^2}\Biggr)
  \frac{e}{\pi} A_\nu(\omega,k)
\end{equation}
with $q=(\omega,k)$ and $q^2=\omega^2-k^2$, from which one can easily
find that
\begin{equation}
 j_V^1(\omega,k) = -\frac{\omega^2}{\omega^2-k^2} \frac{e}{\pi}
  A^1(\omega,k) \;.
\end{equation}
Because $q_W=(e/\pi)\partial^0 A^1$, one can substitute
$A^1=\rmi (2\pi/e)q_W/\omega$ for $A^1$ above, and one can then check
explicitly that the above expression is equivalent to
equation~\eref{eq:jq}.

From the equivalence to the bosonised theory it is very easy to read
the electric current-current fluctuation too.  To this end one should
integrate the $\phi$-field first, and then what remains is a free
massive scalar theory in terms of the $\theta$-field alone.  Then we
trivially get,
\begin{equation}
 \chi_j(x-y) = e^2 \langle j^1(x) j^1(y) \rangle
  = m_\gamma^2\, \partial_0^x
  \partial_0^y \, \langle \theta(x)\theta(y) \rangle \;,
\end{equation}
or in momentum space one can express this as
\begin{equation}
 \chi_j(\omega,k)
  = \frac{m_\gamma^2\, \omega^2}{\omega^2-k^2-m_\gamma^2 +i\epsilon} \;.
\label{eq:chij_full}
\end{equation}
At a first glance this expression looks different from
equation~\eref{eq:chij}.  This is because the above
expression~\eref{eq:chij_full} is a result after resummation of the
bubble-type diagrams, while equation~\eref{eq:chij} is the result of
the one-loop order.  Roughly speaking, $m_\gamma^2$ appears in the
denominator of equation~\eref{eq:chij_full} as a result of infinite
insertion of the polarisation diagram.  This indicates that one can
extract the one-loop result from the leading-order Taylor expansion of
equation~\eref{eq:chij_full} in terms of $m_\gamma^2$.  Such a
procedure actually leads to
\begin{equation}
 \chi_j^{\rm one-loop}(\omega,k) =
  \frac{m_\gamma^2\, \omega^2}{\omega^2-k^2} \;\;\to\;\;
  m_\gamma^2 = \frac{e^2}{\pi} \quad (\mbox{at } k\to 0) \;.
\label{eq:chi-one-loop}
\end{equation}
Therefore,
\begin{eqnarray}
 \chi_j^{\rm one-loop} = \frac{e^2}{\pi}
  &\qquad \mbox{[in (1+1) dimensions]} \nonumber\\
 \longrightarrow\;
  \chi_j^{\rm one-loop} = \frac{|eB|}{2\pi}\cdot \frac{e^2}{\pi} \;,
  &\qquad \mbox{[in (3+1) dimensions]}
\end{eqnarray}
which again coincides with the previous result~\eref{eq:chij}.

In this way the Schwinger model is of great use to understand the
topological properties probed by the magnetic field.  In reference
\cite{Fukushima:2011nu} microscopic calculations to demonstrate how
the dimensional reduction occurs in a way consistent with the momentum
conservation are given with a more detailed result for the
polarisation tensor in (1+1) dimensions embedded in (3+1)-dimensional
gauge fields.


\section{Summary and Outlook}

In this review some of the theoretical approaches to QCD matter in
extreme environments have been picked up.  This direction of physics
is strongly motivated by relativistic heavy-ion collision
experiments.  Furthermore, extreme environments such as the high
temperature, the high baryon density and the strong magnetic field
would enable theorists to attack QCD problems in a treatable way.

Theoretical and experimental researches on finite-$T$ QCD have
achieved the level of the precision science, whereas the
finite-density study of QCD is still controversial.  Theoretical
approaches cannot escape from huge uncertainties, and only the
forthcoming experimental data will be able to impose constraints on
many possibilities proposed from the theoretical side.

New physics opportunities provided by the strong magnetic field
created in the heavy-ion collision are quite intriguing.  A deeper
understanding in this direction would be helpful for the
finite-density study as well.  This is because gluons can be coupled to
both the magnetic field effects and the density effects only through
the quark polarisation processes.

There are many interesting subjects that we had to miss in this
article.  Let us look quickly over some of them.  We did not discuss
the recent developments in the functional approaches to the QCD phase
diagram based on the renormalisation group flow \cite{Braun:2009gm}
and the Dyson-Schwinger equation \cite{Fischer:2009wc,Fischer:2011mz}.
This approach is the most promising among others to attack the problem
of the QCD phase diagram from the first-principle technique.  It has
been understood how quark confinement is realised in terms of the
gluon and ghost propagators \cite{Braun:2007bx}.
Future extensions to the three-flavour
case without uncertainty that stems from the ($T$-dependent) strength
of the U(1)$_{\rm A}$ anomaly would be desirable if possible.
The inhomogeneous chiral condensate (chiral density wave) is the key
concept that may reconcile various states of matter in the baryon-rich
regime;  the QCD critical point, the QCD triple point
\cite{Andronic:2009gj}, Quarkyonic Matter and even the
dimensionally-reduced state in strong magnetic fields
\cite{Basar:2010zd}.  The relation between the chiral density-wave
state and the Polyakov loop dynamics would be a challenging problem
too.

Intense magnetic fields have opened a new direction of physics in the
heavy-ion collision.  The situation realised as a result of the LLL
approximation is similar to that near the Fermi surface at high
density, i.e.\ pseudo-(1+1) dimensionality causes peculiar phenomena
such as superconductivity \cite{Chernodub:2011mc} and sound modes
\cite{Kharzeev:2010gd,Burnier:2011bf} that may have a connection to
the so-called Tomonaga-Luttinger liquid in which no quasi-particle
excitations but only sound modes exist.

All these developments and new possibilities are waiting for further
investigations.

\ack
The author thanks
Oleg Andreev,
Maxim Chernodub,
Gerald Dunne,
Eduardo Fraga,
Tetsuo Hatsuda,
Yoshimasa Hidaka,
Dima Kharzeev,
Youngman Kim,
Toru Kojo,
Larry McLerran,
Shin Nakamura,
Jan Pawlowski,
Rob Pisarski,
Misha Polikarpov,
Marco Ruggieri,
Nan Su,
Harmen Warringa,
Wolfram Weise
for useful discussions.
He also thanks
Jens O.\ Andersen,
Enrique Ruiz Arriola,
Jens Braun,
Christian Fischer,
Igor Shovkovy,
Andrei Smilga,
Naoki Yamamoto
for useful comments.

\section*{References}
\bibliographystyle{utphys}
\bibliography{rhic,lattice,perturb,model,cme,other}

\end{document}